\PassOptionsToPackage{table,dvipsnames}{xcolor}
\documentclass[sigplan,10pt]{acmart}
\renewcommand\footnotetextcopyrightpermission[1]{}
\usepackage{booktabs}
\usepackage{multirow}
\usepackage{amsmath}
\usepackage{algorithm}
\usepackage{algorithmic}
\usepackage{graphicx}
\usepackage{xcolor}
\usepackage{colortbl}
\usepackage{tikz}
\usepackage{pgfmath}
\usepackage{pgfplots}
\pgfplotsset{compat=1.18}
\usepgfplotslibrary{colorbrewer}
\usetikzlibrary{patterns}

\definecolor{cpurow}{HTML}{EBF5FB}    % light blue for CPU rows
\definecolor{gpurow}{HTML}{FDEDEC}    % light red/pink for GPU rows
\definecolor{bestcell}{HTML}{27AE60}  % green for best values
\definecolor{headerrow}{HTML}{2C3E50} % dark header
\definecolor{secheader}{HTML}{F0F0F0} % section header grey
\newcommand{\best}[1]{\textbf{\textcolor{bestcell}{#1}}}
\newcommand{\cpubg}{\rowcolor{cpurow}}
\newcommand{\gpubg}{\rowcolor{gpurow}}
\AtBeginDocument{%
  }

\setcopyright{none}
\copyrightyear{2026}
\acmYear{2026}
\acmDOI{}
\acmConference[ATC '26]{USENIX Annual Technical Conference}{July 2026}{Boston, MA, USA}
\acmISBN{}

\begin{document}

\title{Do You Really Need a GPU to Guard Your LLM? CPU-Class Classifiers and Multi-Stage Pipelines for Safety Enforcement at Scale}

\author{Vasudev Majhi}
\affiliation{%
  \institution{BITS Pilani}
  \city{Pilani}
  \country{India}
}
\email{f20212750@pilani.bits-pilani.ac.in}

\author{Dhruv Gupta}
\affiliation{%
  \institution{BITS Pilani}
  \city{Pilani}
  \country{India}
}
\email{f20221215@pilani.bits-pilani.ac.in}

\author{Advait Singh}
\affiliation{%
  \institution{BITS Pilani}
  \city{Pilani}
  \country{India}
}
\email{F20220636@pilani.bits-pilani.ac.in}

\author{Matthew Barker}
\affiliation{%
  \institution{Trustwise}
  \city{London}
  \country{United Kingdom}
}
\email{matthew@trustwise.ai}

\author{Dhruv Kumar}
\affiliation{%
  \institution{BITS Pilani and Trustwise}
  \city{Pilani}
  \country{India}
}
\email{dhruv.kumar@pilani.bits-pilani.ac.in}

\renewcommand{\shortauthors}{Majhi et al.}

\begin{abstract}
Safety classifiers that screen LLM inputs for jailbreak attempts have become standard deployment components, yet almost all production systems rely on GPU-based models: fine-tuned transformers and LLM-as-a-judge pipelines. These approaches impose significant per-query latency and infrastructure cost. Very little research has asked whether CPU-based classifiers, such as support vector machines and gradient-boosted trees trained on TF-IDF features, can match their accuracy across the conditions that production deployments encounter.

We evaluate five CPU classifier families, Mamba-130M as an SSM-based GPU classifier, and transformer-based GPU models (DeBERTa-v3 and Gemma-2B with LoRA) across nine jailbreak sources and three regimes: in-distribution (D1), out-of-distribution (D2), and adversarially obfuscated (D3). On D1, the best CPU classifier matches the best transformer GPU model at roughly one-fifth the deployment cost. On D2, CPU classifiers fail via confident miscalibration, producing high-confidence false negatives that bypass escalation entirely. On D3, CPU classifiers outperform transformer GPU models by more than 26 percentage points in F1.

Based on these complementary failure modes, we design \textsc{GuardChain}, a three-stage safety pipeline (Regex $\to$ CPU $\to$ GPU) that routes each prompt to the cheapest stage capable of a confident decision. The CPU stage alone resolves 80\% of in-distribution prompts at near-peak accuracy, and the GPU stage recovers the out-of-distribution failures. For practitioners deploying LLM safety at scale, this work provides evidence that GPU-class infrastructure is unnecessary for the majority of traffic.
\end{abstract}




\maketitle
\pagestyle{plain}

%% ============================================================
\section{Introduction}
\label{sec:intro}

Large language models (LLMs) power conversational assistants, coding agents, and enterprise search systems serving billions of queries daily~\cite{inan2024llama}.  Every deployed endpoint is an attack surface.  Jailbreak prompts can coerce models into generating harmful content~\cite{zou2023universal, chao2024jailbreakbench}, and the scale of deployment makes manual review infeasible.  In response, systems such as OpenAI Moderation, NeMo Guardrails, and Llama Guard deploy safety classifiers as a middleware layer that screens each incoming prompt before it reaches the underlying model~\cite{openai2023moderation, nemo2023guardrails, inan2024llama}.

\textbf{The GPU dependency in safety classification.}  Current safety classifiers fall into a broad category of GPU-based model classifiers.  This category includes fine-tuned transformer classifiers, such as DeBERTa-v3 and Gemma-2B adapted for safety detection, and LLM-as-a-judge systems that pass each prompt through a general-purpose model to obtain a safety verdict.  Both require dedicated GPU infrastructure.  Two challenges arise at scale.  First, latency: GPU inference adds tens to hundreds of milliseconds per request, a meaningful penalty in interactive settings.  Second, cost: every prompt that must pass through a GPU guard represents infrastructure that cannot be served on commodity hardware~\cite{strubell2019energy, schwartz2020green}.  In our evaluation, the cheapest fine-tuned transformer guard (DeBERTa-v3 LoRA) costs 45 times more per request than an SVM, and Gemma-2B LoRA costs 80 times more.

\textbf{The research question.}  Can CPU-based classifiers match the accuracy of fine-tuned GPU models while substantially reducing deployment cost?  CPU-based classifiers include traditional machine learning models such as support vector machines (SVMs) and gradient-boosted trees trained on text frequency features (TF-IDF, which weights words by how often they appear relative to the rest of the corpus).  Prior work includes staged guardrail systems but does not provide a systematic comparison of CPU classifiers, compact neural stages, and fine-tuned GPU classifiers across distribution regimes and adversarial perturbations.  This paper addresses that gap.

\textbf{Our work.}  We evaluate five CPU classifier families and two GPU architectural families across nine jailbreak sources and three evaluation regimes.  The in-distribution regime (D1) trains and tests on the same jailbreak source distribution.  The out-of-distribution regime (D2) tests against five held-out jailbreak sources with qualitatively different attack strategies.  The adversarially obfuscated regime (D3) applies leetspeak substitution, character perturbation, and Unicode invisible-character injection to D1 inputs.  The CPU families include three SVM variants (word n-grams, character n-grams, and combined), LightGBM, and Random Forest, all trained on TF-IDF features.  Within the GPU tier, we evaluate two architectural families: Mamba-130M~\cite{gu2024mamba}, a selective state space model (SSM) that processes sequences in linear time via input-dependent recurrence, trained via full curriculum fine-tuning; and DeBERTa-v3 and Gemma-2-2B, transformer-based classifiers fine-tuned with LoRA on the same jailbreak data.  All evaluation is at inference time only; training is outside the scope of this study.

\begin{figure*}[ht]
    \centering
    \begin{tikzpicture}[
        box/.style={draw, rounded corners=3pt, text width=2.4cm, minimum height=1.1cm,
                    align=center, font=\small},
        cpubox/.style={box, fill=blue!7, draw=blue!40},
        gpubox/.style={box, fill=orange!10, draw=orange!50},
        iobox/.style={box, fill=gray!10, draw=gray!50, text width=2.0cm},
        arr/.style={->, thick},
        esc/.style={->, thick, dashed, gray!60},
        lbl/.style={font=\scriptsize\ttfamily},
        stat/.style={font=\small, align=center},
    ]
    \node[iobox]  (inp) at (0,   0) {Incoming\\Prompt};
    \node[cpubox] (s1)  at (3.8, 0) {\textbf{Stage~1}\\\small Regex Filter};
    \node[cpubox] (s2)  at (8.0, 0) {\textbf{Stage~2}\\\small CPU Classifier\\\tiny (SVM / LightGBM)};
    \node[gpubox] (s3)  at (12.5,0) {\textbf{Stage~3}\\\small GPU Model\\\tiny (Mamba-130M)};

    \draw[arr] (inp.east)  -- (s1.west);
    \draw[esc] (s1.east)   -- node[above, lbl]{\texttt{escalate}} (s2.west);
    \draw[esc] (s2.east)   -- node[above, lbl]{\texttt{escalate}} (s3.west);

    \draw[arr] (s1.south) -- ++(0,-0.9) node[below, font=\small\ttfamily]{\texttt{block}/\texttt{allow}};
    \draw[arr] (s2.south) -- ++(0,-0.9) node[below, font=\small\ttfamily]{\texttt{block}/\texttt{allow}};
    \draw[arr] (s3.south) -- ++(0,-0.9) node[below, font=\small\ttfamily]{\texttt{block}/\texttt{allow}};

    \node[stat] at (3.8, -2.7) {$<$0.2\,ms};
    \node[stat] at (8.0, -2.7) {3--50\,ms \\ \$0.17--2.77/1M};
    \node[stat] at (12.5,-2.7) {24.3\,ms \\ \$6.80/1M};
    \end{tikzpicture}
    \caption{Overview of the \textsc{GuardChain} three-stage middleware pipeline.  Each incoming prompt passes through Stage~1 (regex filter), Stage~2 (CPU classifier, SVM or LightGBM), and Stage~3 (GPU model, Mamba-130M), with explicit \texttt{block}/\texttt{allow}/\texttt{escalate} decisions at each stage.  Prompts resolved at early stages bypass later stages, reducing overall latency and cost.  Per-stage latency and approximate cost are shown below each box.}
    \label{fig:pipeline_overview}
    \vspace{-2mm}
\end{figure*}

\textbf{Main results.}  On D1, CPU classifiers match transformer-based GPU models at a fraction of the cost: LightGBM matches Gemma-2B LoRA to within 1 percentage point in F1 at roughly one-fifth the deployment cost.  On OOD data (D2), all CPU classifiers fall below F1\,=\,0.43.  The failure is not ordinary uncertainty.  TF-IDF assigns near-zero weight to vocabulary absent from the training sources, and 72\% of OOD attacks receive confident benign predictions that never reach the cascade's GPU stage — a failure mode we call confident miscalibration.  On adversarially obfuscated inputs (D3), the pattern reverses: LightGBM outperforms Gemma-2B LoRA by more than 26 percentage points in F1.  Mamba-130M handles all three regimes, achieving F1\,=\,0.960 on D1 and 0.927 on D3, and discriminating OOD attacks where CPU classifiers confidently fail.  The 23\% overlap in SVM and Mamba-130M errors on D1 (Figure~\ref{fig:venn}) shows that the two families fail on largely disjoint inputs.  This empirical complementarity aligns with prior cascading and routing systems~\cite{teerapittayanon2016branchynet, chen2023frugalgpt, ong2025routellm}, motivating a composable pipeline.  In \textsc{GuardChain}, the CPU stage resolves 80\% of in-distribution traffic at near-peak accuracy, and routing 20\% of OOD inputs to the GPU stage lifts pipeline F1 from 0.435 to 0.573 on D2.  Taken together, these results show that GPU-class guard infrastructure is not required for the majority of traffic.

\textbf{Research contributions.}  The contributions of this paper are:

\begin{enumerate}
    \item A systematic evaluation of five CPU classifier families and two GPU architectural families (SSM-based and transformer-based) for jailbreak detection across nine sources and three regimes.  To our knowledge, this is the first evaluation of LightGBM and Mamba selective state space models as safety classifiers.

    \item A characterisation of confident miscalibration as a failure mode distinct from ordinary uncertainty: on OOD data, TF-IDF classifiers produce high-confidence false negatives that bypass cascade escalation, rather than uncertain predictions that would trigger it.  This failure mode requires explicit architectural mitigation, not just threshold tuning.

    \item The design and comprehensive evaluation of \textsc{GuardChain}, a three-stage safety pipeline (Regex $\to$ CPU $\to$ GPU) with five variants evaluated across all three regimes.  We show that the CPU stage handles 80--95\% of in-distribution and obfuscated traffic at near-peak accuracy, that the GPU stage is critical for OOD inputs where CPU classifiers fail silently, and that the full pipeline reduces deployment cost by up to 5$\times$ relative to running the GPU stage on every prompt.
\end{enumerate}

%% ============================================================
\section{Background and Related Work}
\label{sec:related}

We organise prior work along two threads.  The first thread (Sections~\ref{sec:rel_threats}--\ref{sec:rel_cpu}) traces the evolution of LLM safety classifiers from GPU-heavy guard models toward increasingly efficient CPU-class designs, and identifies where this paper sits in that progression.  The second thread (Sections~\ref{sec:rel_ssm}--\ref{sec:rel_cascades}) covers the two architectural ingredients of \textsc{GuardChain}: selective state space models and cascade and routing pipelines.

\subsection{LLM Safety Threats and Evaluation Benchmarks}
\label{sec:rel_threats}

The proliferation of LLMs in production has catalysed substantial research on automated safety mechanisms.  Jahan and Oussalah~\cite{jahan2023review} provide a PRISMA review of hate speech detection and highlight that most systems are evaluated on in-distribution splits.  Zou et al.~\cite{zou2023universal} demonstrate universal adversarial suffixes.  Chao et al.~\cite{chao2024jailbreakbench} introduce JailbreakBench and Mazeika et al.~\cite{mazeika2024harmbench} introduce HarmBench.  Chen et al.~\cite{chen2025llm} propose perplexity-based defences, complementing earlier work that detects adversarial prompts via statistical signals such as perplexity and anomaly detection~\cite{alon2023detecting}.  Safety frameworks such as the OWASP LLM Top 10 identify these as critical risks in deployed systems~\cite{owasp2023llm}, and large-scale evaluation frameworks characterise the broader capability and cost trade-off~\cite{liang2022helm, srivastava2022bbh}.  Bassani et al.~\cite{bassani2024guardbench} aggregate many safety datasets into a standardised evaluation pipeline across guard models.  That benchmark does not include OOD source-partitioned splits or adversarial obfuscation augmentation, which motivates our additional corpus construction (Section~\ref{sec:datasets}).

Modern NLP models are also known to rely on superficial patterns and degrade under distribution shift~\cite{geirhos2020shortcut, hendrycks2019robustness, koh2021wilds, ribeiro2020checklist, mccoy2019right}.  Recent malicious-prompt classifier work similarly shows that standard train/test splits can overestimate performance relative to leave-one-dataset-out evaluation~\cite{fomin2026benchmarks}.  In contrast to prior safety evaluations, which mostly report in-distribution results, we evaluate classifiers across both in-distribution and OOD regimes, providing a systematic characterisation of how CPU and GPU classifiers compare under distribution shift.

\subsection{GPU-Based Safety Models}
\label{sec:rel_gpu}

GPU-based guard models have dominated the recent literature.  Inan et al.~\cite{inan2024llama} introduce Llama Guard for input and output safety filtering, and platforms such as NeMo Guardrails~\cite{nemo2023guardrails} deploy GPU-based models for runtime enforcement.  LLM-as-a-judge approaches achieve high accuracy at significant computational cost~\cite{perez2022red}.  Recent work has broadened coverage across risk categories and model families.  Han et al.~\cite{han2024wildguard} introduce WildGuard, a guard model covering a broad set of risk categories.  Zeng et al.~\cite{zeng2024shieldgemma} introduce ShieldGemma, a safety classifier series spanning several model scales.  Ghosh et al.~\cite{ghosh2024aegis} propose AEGIS, an adaptive ensemble.  Li et al.~\cite{li2024salad} provide the SALAD-Bench benchmark and the MD-Judge safety classifier, and Zhang et al.~\cite{zhang2024shieldlm} offer customisable, explainable bilingual safety detection.  These models all assume that GPU inference is necessary for accurate safety classification.  Our evaluation includes DeBERTa-v3 and Gemma-2B as representative fine-tuned GPU baselines (Appendix~\ref{app:finetuned}) and Granite-Guardian-2B as an off-the-shelf baseline (Appendix~\ref{app:offshelf}).  Studies of guard-model calibration under jailbreak attacks~\cite{guardcalib2024} are directly relevant to the confident miscalibration phenomenon we characterise in Section~\ref{sec:gc_d2}.

\subsection{Lightweight and CPU-Deployable Safety Classifiers}
\label{sec:rel_cpu}

Classical machine learning remains relevant for low-latency filtering.  Joachims~\cite{joachims1998svm} demonstrates the effectiveness of SVM-based text classification.  Lin et al.~\cite{lin2023linear} show that linear classifiers are competitive baselines.  Prior work also explores hate speech detection using neural models~\cite{macavaney2019hatespeech, kumar2024hybrid, khan2022bichat, chen2024hate}, and domain-specific pretrained models such as HateBERT improve performance on abusive-language detection~\cite{caselli2021hatebert}.  Model compression and distillation approaches such as DistilBERT and TinyBERT demonstrate that lightweight models can be competitive at reduced cost~\cite{sanh2019distilbert, jiao2020tinybert}.  More recent work directly targets CPU deployment of guard models.  Meta's PromptGuard~\cite{meta2024promptguard} is a compact multi-label classifier explicitly designed for CPU deployment.  Faillenot et al.~\cite{faillenot2024llamaguardquant} show that a quantised and pruned Llama Guard achieves real-time inference on mobile CPUs.  A further approach trains penalised logistic regressions on the intermediate hidden states of pruned LLMs for safety classification~\cite{lec2024lightweight}.  JavelinGuard explores compact transformer classifiers for LLM security that are intended for CPU-class deployment~\cite{javelinguard2025}.  Three recent empirical studies are particularly close to our first contribution.  PromptScreen uses text normalisation, TF-IDF features, and a Linear SVM inside a multi-stage jailbreak mitigation pipeline~\cite{promptscreen2026}; our work generalises this direction beyond a single SVM stage by comparing multiple CPU families, obfuscated and OOD regimes, and full GuardChain variants.  A second study compares TF-IDF with logistic regression against fine-tuned BERT for jailbreak detection~\cite{anon2024noveljailbreaks}.  Galinkin and Sablotny~\cite{galinkin2024improved} combine pretrained text embeddings with classical machine learning classifiers for jailbreak detection; this embedding-based line is complementary to our TF-IDF analysis because it uses pretrained representations rather than surface lexical features, which is exactly where our OOD failure analysis finds TF-IDF to be brittle.

Compared to this body of work, the missing piece is a systematic head-to-head comparison of multiple CPU classifier families, including gradient-boosted trees, against compact neural and fine-tuned GPU guard models across multiple attack vectors and distribution regimes.  Our paper supplies this comparison and, to the best of our knowledge, provides the first evaluation of LightGBM and Mamba selective state space models as safety classifiers.

\subsection{State Space Models}
\label{sec:rel_ssm}

Gu and Dao~\cite{gu2024mamba} introduce Mamba, a selective state space model achieving linear-time inference through input-dependent gating, building on earlier structured state space work~\cite{gu2021s4}.  Mamba has since been evaluated on long text classification~\cite{mamba2024longtext} and as a text reranker~\cite{ssm2024rerankers}, which suggests that its hidden state captures semantic structure useful for classification beyond next-token prediction.  To the best of our knowledge, no prior work has evaluated state space models as safety classifiers.  Mamba's linear-time architecture and small 130M-parameter footprint position it as a middle ground between surface-feature CPU classifiers and larger transformer guard models, which directly motivates its role as the GPU-assisted Stage~3 in \textsc{GuardChain}.  Appendix~\ref{app:mamba_latency} separately reports true CPU-only latency, which is substantially slower in our environment.

\subsection{Cascade and Routing Architectures}
\label{sec:rel_cascades}

The idea of routing inputs through models of increasing cost predates modern LLMs, as in BranchyNet~\cite{teerapittayanon2016branchynet} and Mixture of Experts~\cite{shazeer2017moe}.  More directly relevant to \textsc{GuardChain}, Chen et al.~\cite{chen2023frugalgpt} propose FrugalGPT, which cascades increasingly capable LLMs based on quality estimates to reduce serving cost.  Ding et al.~\cite{ding2024hybridllm} train a difficulty router that reduces expensive-model calls without quality loss.  Ong et al.~\cite{ong2025routellm} learn LLM routing from preference data.  In the safety domain, SafeRoute trains a binary router that sends hard prompts to a larger guard model while allowing easier prompts to use a smaller guard~\cite{saferoute2025}; GuardChain differs by routing from classical CPU classifiers to a compact GPU stage in a three-stage pipeline, and by evaluating the resulting cascade under OOD and obfuscated regimes.  Chennabasappa et al.~\cite{chennabasappa2025llamafirewall} propose LlamaFirewall, which composes a prompt-injection guard, a chain-of-thought auditor, and a static-analysis stage.  LlamaFirewall is structurally similar to \textsc{GuardChain} but operates at production scale without a systematic CPU-versus-GPU classifier comparison or OOD characterisation.  Our contribution is to take these cascade principles, apply them to safety classification, and characterise where each stage adds value across in-distribution, OOD, and obfuscated regimes, surfacing failure modes such as confident miscalibration that are specific to the safety setting.

%% ============================================================
%% ============================================================
\section{Evaluation Setup}
\label{sec:setup}

\subsection{Datasets}
\label{sec:datasets}

We evaluate jailbreak detection across three distribution regimes.  Table~\ref{tab:datasets} summarises the corpora.  Table~\ref{tab:datasources} lists all nine source datasets and their OOD partition assignment.  Hate speech, toxicity, and PII datasets used in supplemental evaluation are described in Appendix~\ref{app:multitask}.

We aggregate nine public sources (Jailbreak Classification, GPT-Fuzzer, InTheWild, JailBreakV-28K, RedTeam-2K, AdvBench, JBB Behaviors, HarmBench, StrongReject), totalling 43{,}523 deduplicated samples (55\% malicious).  We prepare two versions.  The \textbf{in-distribution split (D1)} shuffles all sources and partitions into 70/15/15 train/val/test with zero text-overlap leakage (verified by MD5-hash set intersection; Appendix~\ref{app:datasets}).  The \textbf{out-of-distribution split (D2)} partitions by source: training uses four sources (Jailbreak Classification, GPT-Fuzzer, InTheWild, JailBreakV-28K), and the test set uses five entirely held-out sources (RedTeam-2K, AdvBench, JBB Behaviors, HarmBench, StrongReject) whose attack templates differ qualitatively in phrasing and structure from the training sources.  We additionally generate an \textbf{obfuscated variant (D3)} by applying three stochastic transformations to D1 inputs.  Let $\mathcal{T} = \mathcal{T}_{\text{leet}} \circ \mathcal{T}_{\text{perturb}} \circ \mathcal{T}_{\text{unicode}}$ denote their composition.

\textbf{(1) Leetspeak substitution} ($\mathcal{T}_{\text{leet}}$) applies a stochastic character mapping with probability $p_{\text{leet}} = 0.3$ per character:
\begin{equation}
    \phi(c) = \begin{cases}
        \{4, @\} & c = \texttt{a} \\
        \{3\} & c = \texttt{e} \\
        \{0\} & c = \texttt{o} \\
        \{5, \$\} & c = \texttt{s} \\
        \{c\} & \text{otherwise}
    \end{cases}
    \label{eq:leet}
\end{equation}

\textbf{(2) Character perturbation} ($\mathcal{T}_{\text{perturb}}$) applies one of four strategies uniformly at random per word with probability $p_{\text{perturb}} = 0.15$: (i)~swap adjacent characters, (ii)~delete a character, (iii)~insert a random character, or (iv)~substitute a vowel with a random vowel.

\textbf{(3) Unicode injection} ($\mathcal{T}_{\text{unicode}}$) inserts invisible characters at random positions: zero-width spaces (U+200B), control characters (U+0000--U+001F), and non-breaking spaces (U+00A0) at a density of one invisible character per 50 visible characters on average.

\begin{table}[t]
  \caption{Jailbreak dataset sources and raw sample counts.  All nine sources contribute to D1 (randomly shuffled across sources, 70/15/15 train/val/test split).  The D2 Role column shows how each source is assigned in the out-of-distribution evaluation: rows 1--4 form the D2 training partition, and rows 5--9 are held-out D2 test data only.}
  \label{tab:datasources}
  \small
  \begin{tabular}{lrrl}
    \toprule
    \rowcolor{headerrow}\textcolor{white}{\textbf{Dataset}} & \textcolor{white}{\textbf{Samples}} & \textcolor{white}{\textbf{Host}} & \textcolor{white}{\textbf{D2 Role}} \\
    \midrule
    Jailbreak Classification & 1{,}998 & HuggingFace & D2 train \\
    \rowcolor{gray!5} GPT-Fuzzer & 7{,}700 & GitHub & D2 train \\
    InTheWild (Combined) & 15{,}140 & GitHub & D2 train \\
    \rowcolor{gray!5} JailBreakV-28K & 28{,}000 & HuggingFace & D2 train \\
    RedTeam-2K & 2{,}000 & HuggingFace & D2 test \\
    \rowcolor{gray!5} AdvBench & 520 & GitHub & D2 test \\
    JBB Behaviors & 200 & HuggingFace & D2 test \\
    \rowcolor{gray!5} HarmBench & 400 & GitHub & D2 test \\
    StrongReject & 313 & GitHub & D2 test \\
    \midrule
    \rowcolor{secheader}\textbf{Total (before dedup.)} & \textbf{56{,}271} & & \\
    \bottomrule
  \end{tabular}
\end{table}

\begin{table}[t]
  \caption{Dataset statistics after preprocessing and deduplication.  Per-task difficulty is analysed in Appendix~\ref{app:easiness}.}
  \label{tab:datasets}
  \centering
  \begin{tabular}{lrrl}
    \toprule
    \rowcolor{headerrow}\textcolor{white}{\textbf{Dataset}} & \textcolor{white}{\textbf{Train / Test}} & \textcolor{white}{\textbf{\% Pos.}} & \textcolor{white}{\textbf{Task}} \\
    \midrule
    \rowcolor{secheader}\multicolumn{4}{l}{\textit{Complete (in-distribution)}} \\
    Jailbreak (D1) & 27{,}641 / 5{,}924 & 55\% & Jailbreak \\
    \rowcolor{gray!5} Obfuscated (D3) & 27{,}641 / 5{,}924 & 55\% & Jailbreak \\
    \midrule
    \rowcolor{secheader}\multicolumn{4}{l}{\textit{Partitioned (out-of-distribution)}} \\
    Jailbreak OOD (D2) & 30{,}763 / 3{,}296 & 97\%$^*$ & Jailbreak \\
    \bottomrule
    \multicolumn{4}{l}{\footnotesize $^*$Test set heavily skewed toward the jailbreak class.}
  \end{tabular}
\end{table}

These regimes are constructed from the same underlying jailbreak corpus, but they answer different generalisation questions.  D1 is the standard in-distribution setting: examples from all nine source families are mixed and then randomly partitioned into train, validation, and test splits.  D2 reuses the same cleaned corpus but changes the partitioning rule from random splitting to source-family splitting: four source families are used for training and validation, while five disjoint source families are reserved only for testing.  Thus D2 is an out-of-distribution source-shift evaluation rather than a separate task.  D3 preserves the D1 split sizes and labels but applies adversarial text transformations to the corresponding D1 examples, measuring robustness to surface-form obfuscation.

\subsection{Classifiers}
\label{sec:models}

\textbf{TF-IDF-based classifiers (CPU lexical tier).}  Features are extracted by Term Frequency-Inverse Document Frequency (TF-IDF) over word and character n-grams.  We designed three feature variants spanning different n-gram granularities (Table~\ref{tab:svm_variants_main}).  All three variants are augmented with eight engineered safety-specific features (Table~\ref{tab:eng_features_main}) that capture prompt-level structural cues including length anomalies, punctuation density, upper-case saturation, and character repetition.  We identified these eight signals by systematic inspection of attack patterns across our jailbreak corpus; they target structural regularities common to prompt attacks that TF-IDF alone may miss in short or heavily formatted inputs.

\begin{table}[t]
  \caption{TF-IDF feature variant specifications.}
  \label{tab:svm_variants_main}
  \small
  \begin{tabular}{lccc}
    \toprule
    \rowcolor{headerrow}\textcolor{white}{\textbf{Variant}} & \textcolor{white}{\textbf{Word N-grams}} & \textcolor{white}{\textbf{Char N-grams}} & \textcolor{white}{\textbf{Features}} \\
    \midrule
    \cpubg A & (1,2) & --- & $\sim$15{,}008 \\
    \cpubg B & --- & (3,5) & $\sim$5{,}008 \\
    \cpubg C & (1,2) & (3,5) & $\sim$20{,}008 \\
    \bottomrule
  \end{tabular}
\end{table}

\begin{table}[t]
  \caption{Engineered safety feature definitions ($\mathcal{P}$: punctuation chars; $\mathcal{A}$: alphanumeric chars).}
  \label{tab:eng_features_main}
  \small
  \begin{tabular}{llc}
    \toprule
    \rowcolor{headerrow}\textcolor{white}{\textbf{Feature}} & \textcolor{white}{\textbf{Definition}} & \textcolor{white}{\textbf{Range}} \\
    \midrule
    char\_length & $|d| / \max_{d'} |d'|$ & $[0,1]$ \\
    \rowcolor{gray!5} word\_count & $|\text{words}(d)| / \max_{d'} |\text{words}(d')|$ & $[0,1]$ \\
    avg\_word\_len & $\frac{1}{|\text{words}(d)|}\sum_{w} |w|$ & $\mathbb{R}^+$ \\
    \rowcolor{gray!5} punct\_ratio & $|\{c \in d : c \in \mathcal{P}\}| / |d|$ & $[0,1]$ \\
    upper\_ratio & $|\{c \in d : c \in [A\text{-}Z]\}| / |d|$ & $[0,1]$ \\
    \rowcolor{gray!5} digit\_ratio & $|\{c \in d : c \in [0\text{-}9]\}| / |d|$ & $[0,1]$ \\
    special\_ratio & $|\{c \in d : c \notin \mathcal{A}\}| / |d|$ & $[0,1]$ \\
    \rowcolor{gray!5} repetition & $1 - |\text{unique}(w)| / |\text{words}(d)|$ & $[0,1]$ \\
    \bottomrule
  \end{tabular}
\end{table}

On top of these features we train three model families.  The first is a linear SVM with isotonic calibration~\cite{zadrozny2002transforming}, which produces calibrated posterior probabilities.  The second is LightGBM~\cite{ke2017lightgbm}, a gradient-boosted decision tree ensemble with class-imbalance correction (\texttt{scale\_pos\_weight} $= n_-/n_+$) and F1-optimised decision-threshold selection: we sweep $\tau \in [0.05, 0.95]$ on the validation set and choose the threshold maximising positive-class F1, shifting the decision boundary away from the default 0.5 for imbalanced distributions (optimal thresholds per dataset in Appendix~\ref{app:lightgbm}).  The third is a Random Forest~\cite{breiman2001random} with 300 estimators and balanced class weights.

\textbf{Mamba-130M.}  Mamba~\cite{gu2024mamba} is a selective state space model (SSM) that processes a sequence via a discretised linear recurrence:
\begin{equation}
    \mathbf{h}_k = \bar{\mathbf{A}} \cdot \mathbf{h}_{k-1} + \bar{\mathbf{B}}_k \cdot x_k, \quad y_k = \mathbf{C}_k \cdot \mathbf{h}_k
    \label{eq:ssm_main}
\end{equation}
The discretised matrices $\bar{\mathbf{A}}$ and $\bar{\mathbf{B}}_k$ are derived from continuous parameters via zero-order hold (ZOH) discretisation with step size $\Delta_k$:
\begin{align}
    \bar{\mathbf{A}}_k &= \exp(\Delta_k \mathbf{A}) \label{eq:zoh_main_a} \\
    \bar{\mathbf{B}}_k &= (\Delta_k \mathbf{A})^{-1}(\exp(\Delta_k \mathbf{A}) - \mathbf{I}) \cdot \Delta_k \mathbf{B}_k \label{eq:zoh_main_b}
\end{align}
Mamba's selectivity comes from making $\mathbf{B}_k$, $\mathbf{C}_k$, and $\Delta_k$ \emph{input-dependent} through learned linear projections:
\begin{equation}
    \mathbf{B}_k = \mathbf{W}_B \mathbf{x}_k, \quad \mathbf{C}_k = \mathbf{W}_C \mathbf{x}_k, \quad \Delta_k = \text{softplus}(\mathbf{W}_\Delta \mathbf{x}_k + b_\Delta)
    \label{eq:selective_main}
\end{equation}
where $\mathbf{W}_B, \mathbf{W}_C \in \mathbb{R}^{N \times D}$, $\mathbf{W}_\Delta \in \mathbb{R}^{1 \times D}$.  This selectivity allows the model to accumulate or discard information at each token, which is useful for safety classification: the model can track escalating threat cues distributed across a long prompt without attending to every token pair.  Mamba processes each token in $O(D \cdot N)$ time, giving $O(L)$ total complexity versus $O(L^2 \cdot D)$ for self-attention; full derivation in Appendix~\ref{app:mamba}.  Table~\ref{tab:mamba_arch_main} lists the architecture configuration.

\begin{table}[t]
  \caption{Mamba-130M architecture configuration.}
  \label{tab:mamba_arch_main}
  \small
  \begin{tabular}{lr}
    \toprule
    \rowcolor{headerrow}\textcolor{white}{\textbf{Parameter}} & \textcolor{white}{\textbf{Value}} \\
    \midrule
    Total parameters & 129.1M \\
    \rowcolor{gray!5} Hidden dimension $D$ & 768 \\
    SSM state dimension $N$ & 16 \\
    \rowcolor{gray!5} Number of SSM layers & 24 \\
    Expansion factor & 2 \\
    \rowcolor{gray!5} Conv1d kernel size & 4 \\
    Vocabulary size & 50{,}280 \\
    \rowcolor{gray!5} Max sequence length (Phase~1) & 256 \\
    Max sequence length (Phase~2) & 512 \\
    \bottomrule
  \end{tabular}
\end{table}

We fine-tune Mamba-130M via a two-phase curriculum training strategy designed to give the model broad safety coverage followed by jailbreak specialisation (Table~\ref{tab:curriculum_main}).  Phase~1 trains on hate speech, toxicity, and PII tasks, establishing general safety representations.  Phase~2 trains on jailbreak and adversarially obfuscated inputs, specialising the model on the attack types central to this paper.  The curriculum introduces a coverage trade-off: Phase~2 partially overwrites Phase~1 toxicity representations, a form of catastrophic forgetting analysed quantitatively in Appendix~\ref{app:mamba_checkpoints}.  We select the Phase~2 checkpoint by average F1 across all seven evaluation tasks, choosing \texttt{e3\_end} (Avg F1\,=\,0.786).

\begin{table}[t]
  \caption{Curriculum learning phases for Mamba-130M.}
  \label{tab:curriculum_main}
  \small
  \begin{tabular}{lll}
    \toprule
    \rowcolor{headerrow}\textcolor{white}{\textbf{}} & \textcolor{white}{\textbf{Phase~1}} & \textcolor{white}{\textbf{Phase~2}} \\
    \midrule
    Goal & General safety & Jailbreak specialist \\
    \rowcolor{gray!5} Datasets & Hate + PII + Toxic & Jailbreak + Obfusc. \\
    Samples & $\sim$500K & $\sim$510K \\
    \rowcolor{gray!5} Epochs & 3 & 5 \\
    Max length & 256 tokens & 512 tokens \\
    \rowcolor{gray!5} Learning rate & 2e-5 & 1e-5 \\
    Checkpoints & 6 (per $\frac{1}{2}$ epoch) & 12 (per $\frac{1}{2}$ epoch) \\
    \bottomrule
  \end{tabular}
\end{table}

\textsc{Mamba-GPU} (the fine-tuned checkpoint running on GPU) serves as Stage~3 in \textsc{GuardChain}.  To our knowledge, this is the first evaluation of a selective state space model for safety classification.  CPU-only inference for this checkpoint reaches 1{,}696\,ms per request (Appendix~\ref{app:mamba_latency}), making it impractical as a CPU classifier.

\textbf{Fine-tuned GPU baselines.}  To compare against neural guard models trained on the target distribution, we fine-tune two GPU models on the same jailbreak training data using one fixed LoRA configuration (rank=8, $\alpha$=16, lr=2e-4, 3 epochs): Gemma-2-2b and DeBERTa-v3-base.  The fine-tuning configuration (bf16, per-architecture \texttt{target\_modules}, explicit \texttt{task\_type=SEQ\_CLS}) is described in Appendix~\ref{app:lora_corrections}.  Off-the-shelf GPU guard-model evaluations (PromptGuard-86M, Llama-Guard-3-1B, ShieldGemma-2B, WildGuard-7B) are provided in Appendix~\ref{app:offshelf} for reference.

\subsection{Evaluation Protocol}
\label{sec:protocol}

\begin{sloppypar}
We evaluate across three distribution regimes.  We use $A\!\rightarrow\!B$ notation to make the train/test source explicit: the left-hand side denotes the training regime used for fitting the model and selecting thresholds/checkpoints, and the right-hand side denotes the held-out test regime used for final reporting.  Thus D1$\rightarrow$D1 trains on the random mixed-source D1 training partition and tests on the D1 test partition; D2$\rightarrow$D2 trains on the D2 source-partitioned training sources and tests on the held-out D2 source families; D1$\rightarrow$D2 trains only on the D1 random mixed-source training partition and evaluates on the held-out D2 source-family test partition, with no use of the D2 training partition; and D1$\rightarrow$D3 trains on D1 and tests on the obfuscated D3 test partition.  Multi-task evaluation (hate speech, toxicity, PII) is in Appendix~\ref{app:multitask}.
\end{sloppypar}

The primary metric is F1 on the positive (unsafe) class; we also report accuracy, precision, and recall.  Latency is single-request wall-clock time (batch=1, mean over the full test set, models pre-loaded).

CPU inference is conducted on a 32-core server with 503\,GB RAM (CUDA disabled for all CPU rows).  GPU inference latencies are measured on NVIDIA A100 80\,GB PCIe nodes.  AWS on-demand pricing is used as a reference for deployment cost estimates.  Full environment details are in Appendix~\ref{app:environment}.

%% ============================================================
%% ============================================================
\section{Model Exploration}
\label{sec:exploration}

Table~\ref{tab:exploration} presents results for all classifiers across the three evaluation regimes, including accuracy, precision, recall, and F1 for each dataset.

\begin{table*}[ht]
  \caption{Jailbreak classification results across three evaluation regimes.  Accuracy, precision, recall, and F1 are reported for all three datasets.  Latency is single-request wall-clock time (CPU rows: 32-core server; GPU rows: A100 80\,GB).  $^\ddagger$D2 test set is 97\% malicious; an always-positive classifier achieves F1\,=\,0.985.}
  \label{tab:exploration}
  \small
  \setlength{\tabcolsep}{2pt}
  \begin{tabular}{p{2.4cm}l*{13}{r}}
    \toprule
    \rowcolor{headerrow}
      \multicolumn{2}{c}{} &
      \multicolumn{4}{c}{\textcolor{white}{\textbf{D1 (in-distribution)}}} &
      \multicolumn{4}{c}{\textcolor{white}{\textbf{D2 (OOD $^\ddagger$)}}} &
      \multicolumn{4}{c}{\textcolor{white}{\textbf{D3 (obfuscated)}}} &
      \multicolumn{1}{c}{} \\
    \cmidrule(lr){3-6}\cmidrule(lr){7-10}\cmidrule(lr){11-14}
    \rowcolor{headerrow}
      \textcolor{white}{\textbf{Model}} & \textcolor{white}{\textbf{HW}} &
      \textcolor{white}{\textbf{Acc}} & \textcolor{white}{\textbf{Prec}} & \textcolor{white}{\textbf{Rec}} & \textcolor{white}{\textbf{F1}} &
      \textcolor{white}{\textbf{Acc}} & \textcolor{white}{\textbf{Prec}} & \textcolor{white}{\textbf{Rec}} & \textcolor{white}{\textbf{F1}} &
      \textcolor{white}{\textbf{Acc}} & \textcolor{white}{\textbf{Prec}} & \textcolor{white}{\textbf{Rec}} & \textcolor{white}{\textbf{F1}} &
      \textcolor{white}{\textbf{Lat.}} \\
    \midrule
    \rowcolor{secheader}\multicolumn{15}{l}{\textit{CPU (TF-IDF)}} \\
    \cpubg SVM (word n-gram)  & CPU & 0.962 & 0.974 & 0.955 & 0.964 & 0.290 & 0.992 & 0.270 & 0.424 & 0.903 & 0.908 & 0.912 & 0.910 & \best{3.4} \\
    \cpubg SVM (char n-gram)  & CPU & 0.956 & 0.968 & 0.950 & 0.959 & 0.263 & 0.999 & 0.241 & 0.388 & 0.928 & 0.936 & 0.930 & 0.933 & 4.6 \\
    \cpubg SVM (word+char)    & CPU & 0.961 & 0.970 & 0.958 & 0.964 & 0.256 & 0.996 & 0.234 & 0.379 & 0.928 & 0.932 & 0.933 & 0.932 & 5.3 \\
    \cpubg LightGBM           & CPU & 0.966 & 0.978 & 0.959 & 0.968 & 0.281 & 0.996 & 0.260 & 0.412 & \best{0.945} & \best{0.955} & 0.942 & \best{0.948} & 49.4 \\
    \cpubg Random Forest      & CPU & 0.958 & 0.978 & 0.941 & 0.960 & 0.089 & \best{1.000} & 0.060 & 0.114 & 0.935 & 0.933 & \best{0.946} & 0.939 & 67.2 \\
    \midrule
    \rowcolor{secheader}\multicolumn{15}{l}{\textit{GPU: SSM}} \\
    \gpubg Mamba-130M         & GPU & 0.958 & 0.964 & 0.956 & 0.960 & 0.963 & 0.987 & 0.975 & \best{0.981}$^\ddagger$ & 0.921 & 0.913 & 0.942 & 0.927 & 24.3 \\
    \midrule
    \rowcolor{secheader}\multicolumn{15}{l}{\textit{GPU: Transformer (LoRA)}} \\
    \gpubg DeBERTa-v3         & GPU & 0.961 & 0.971 & 0.955 & 0.963 & 0.489 & 0.988 & 0.479 & 0.645 & 0.573 & 0.567 & 0.862 & 0.684 & 27.5 \\
    \gpubg Gemma-2-2B         & GPU & \best{0.973} & \best{0.978} & \best{0.971} & \best{0.974} & 0.668 & 0.990 & 0.665 & 0.795 & 0.729 & 0.896 & 0.557 & 0.687 & 48.5 \\
    \bottomrule
  \end{tabular}
\end{table*}

\subsection{In-Distribution Performance (D1)}
\label{sec:exp_d1}

CPU classifiers match transformer-based GPU models on in-distribution jailbreak detection.  LightGBM achieves F1\,=\,0.968, within 1 percentage point of Gemma-2B LoRA (F1\,=\,0.974).  This gap makes the choice between them a question of deployment cost, not peak accuracy: LightGBM costs \$2.77 per 1\,M requests versus \$13.54 for Gemma-2B LoRA, a 4.9$\times$ cost reduction.  SVM, the fastest option at 3.4\,ms and \$0.17/1M, matches LightGBM's D1 F1 at 80$\times$ lower cost than Gemma-2B LoRA.

SVM classifiers achieve F1 of 0.959--0.964 at 3.4--5.3\,ms and precision of 0.968--0.974, the fastest and cheapest options across all classifiers.  Gemma-2B LoRA reaches F1\,=\,0.974 and precision 0.978 at 48.5\,ms, but runs 14$\times$ slower than SVM and at 80$\times$ higher cost.  The three SVM variants reach comparable D1 accuracy, confirming that the choice of n-gram type matters less on in-distribution data than under distribution shift.

Mamba-130M achieves F1\,=\,0.960 and precision\,=\,0.964 at 24.3\,ms, competitive with all CPU classifiers on D1 accuracy while running at 7$\times$ lower latency than LightGBM (49.4\,ms) and at roughly one-fifteenth of Gemma-2B's parameter count.

\subsection{Out-of-Distribution Robustness (D2)}
\label{sec:exp_d2}

The D2 test set is 97\% malicious by construction (3,196 malicious, 100 benign), so F1 alone is not a sufficient metric here: an always-positive classifier already achieves F1\,=\,0.985.  The meaningful question is whether classifiers fail silently via confident false negatives or whether they remain discriminative.

CPU classifiers fail silently.  All SVM variants fall below F1\,=\,0.43, and Random Forest reaches F1\,=\,0.114.  The failure is not ordinary uncertainty.  TF-IDF assigns near-zero weight to vocabulary absent from the training sources, producing high-confidence benign predictions on novel attack strategies.  We call this confident miscalibration: 72.2\% of OOD attacks receive high-confidence benign predictions rather than uncertain ones that would warrant further review (Appendix~\ref{app:ood_confusion}).

Mamba-130M achieves F1\,=\,0.981 on the full D2 test at 98.7\% precision and 97.5\% recall, maintaining genuine discrimination rather than defaulting to the always-positive strategy.  Its learned contextual representations transfer to unseen jailbreak strategies because they capture semantic attack intent rather than surface lexical form.  Transformer-based GPU models (DeBERTa: F1\,=\,0.645, Gemma-2B: F1\,=\,0.795) outperform CPU classifiers on D2 but fall substantially below Mamba-130M, which captures semantic intent rather than source-specific surface patterns.

\subsection{Obfuscation Robustness (D3)}
\label{sec:exp_d3}

CPU classifiers hold up well on D3.  LightGBM achieves F1\,=\,0.948 and char-SVM achieves F1\,=\,0.933.

Transformer-based GPU models degrade sharply: Gemma-2B drops to F1\,=\,0.687 and DeBERTa to 0.684.  LightGBM outperforms Gemma-2B LoRA by 26.1 percentage points in F1 on D3 (0.948 vs 0.687) while costing 4.9$\times$ less at deployment.  Mamba-130M achieves F1\,=\,0.927 on D3, a drop of 3.3 percentage points from D1 (0.960), benefiting from Phase~2 curriculum training that included obfuscated examples.

\subsection{Summary: Complementary Failure Modes}
\label{sec:exp_summary}

No single classifier wins across all three regimes.  CPU classifiers lead on D3, Mamba-130M handles D2 discriminatively where CPU classifiers fail, and both are comparable on D1.  The 23\% overlap in SVM and Mamba-130M misclassifications on D1 (Figure~\ref{fig:venn}) confirms they fail on disjoint inputs.  This motivates \textsc{GuardChain}: a composable pipeline where each stage handles the inputs it is most suited for and escalates the rest.

%% ============================================================
%% ============================================================
\section{\textsc{GuardChain}}
\label{sec:guardchain}

\subsection{Design and Pipeline Variants}
\label{sec:pipeline_design}

\begin{figure}[t]
    \centering
    \includegraphics[width=0.42\textwidth]{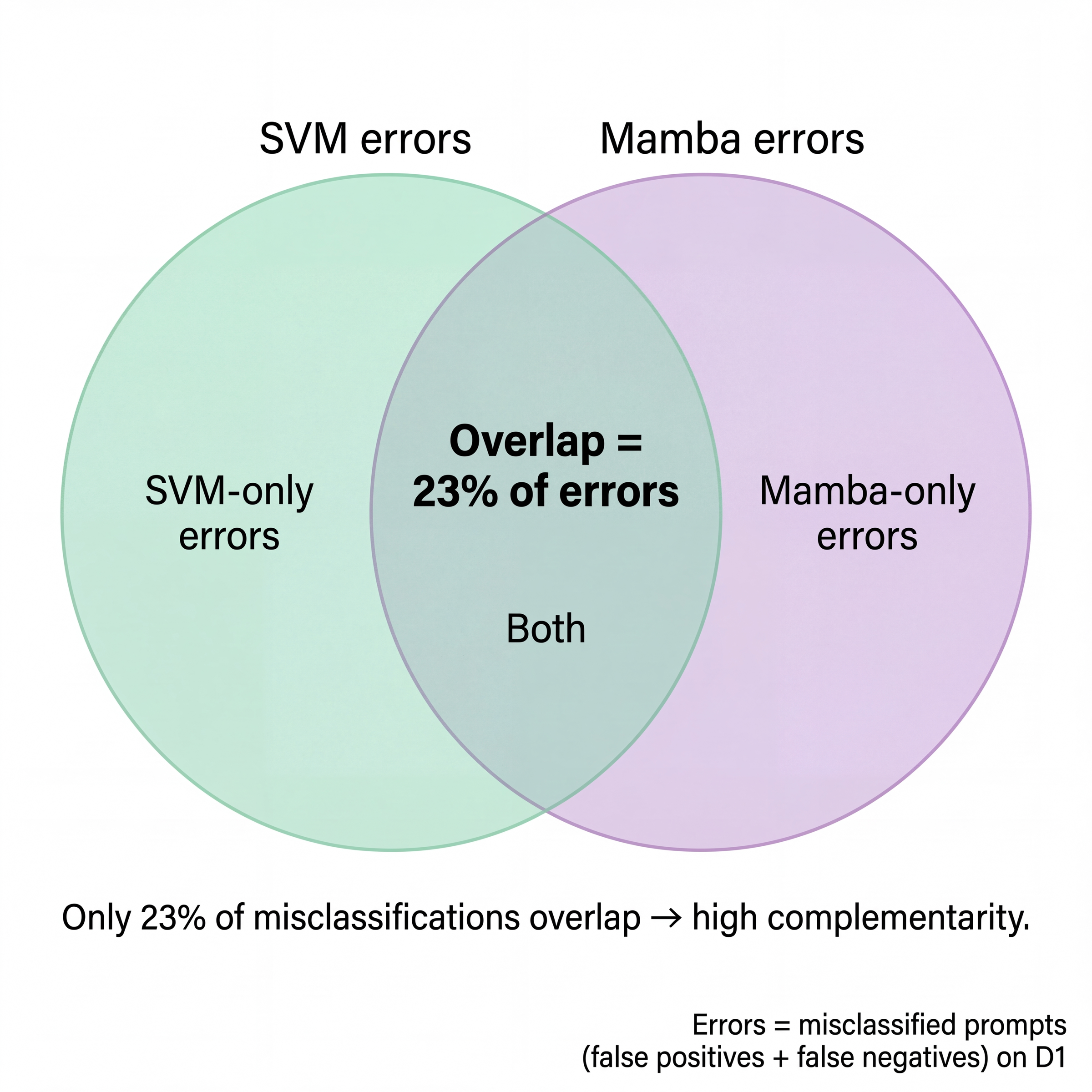}
    \caption{Venn diagram of misclassified prompts by SVM and \textsc{Mamba-GPU} on D1.  Only 23\% of errors overlap, confirming that the two model families fail on disjoint inputs and that cascading them yields more than either alone.}
    \label{fig:venn}
    \vspace{-2mm}
\end{figure}

Section~\ref{sec:exploration} establishes that no single classifier is best across all three regimes and that SVM and \textsc{Mamba-GPU} fail on disjoint inputs (Figure~\ref{fig:venn}).  These findings motivate a composable pipeline where each stage handles the inputs it is most suited for and escalates the rest.

\textsc{GuardChain} is a three-stage pipeline, illustrated in Figure~\ref{fig:pipeline_overview}, where each stage implements a common interface returning one of three verdicts: \texttt{block} (reject), \texttt{allow} (pass), or \texttt{escalate} (forward to next stage):

\begin{equation}
    \textsc{Classify}(p) \rightarrow (v, s, m)
    \label{eq:interface}
\end{equation}

\noindent where $p$ is the input prompt, $v \in \{\texttt{block}, \texttt{allow}, \texttt{escalate}\}$ is the verdict, $s \in [0,1]$ is the confidence score, and $m$ is optional metadata.  A prompt is escalated when the current stage confidence is within $\delta = 0.15$ of its decision boundary:

\begin{equation}
    v = \begin{cases}
        \texttt{block} & \text{if } s > 0.5 + \delta \\
        \texttt{allow} & \text{if } s < 0.5 - \delta \\
        \texttt{escalate} & \text{otherwise}
    \end{cases}
    \label{eq:cascade_decision}
\end{equation}

The three stages are: Stage~1, a regex filter covering prompt injection cues and shell injection tokens with sub-millisecond latency; Stage~2, a CPU classifier (LightGBM for the accuracy-optimised variants, SVM for the latency-optimised variant); and Stage~3, a GPU model for prompts Stage~2 cannot confidently resolve.  We evaluate five variants:

\begin{itemize}
    \item \textbf{Baseline-CPU}: LightGBM alone (no regex, no GPU stage).
    \item \textbf{Baseline-GPU}: \textsc{Mamba-GPU} alone.
    \item \textbf{GC-CPU}: Regex $+$ LightGBM (no GPU stage).
    \item \textbf{GC-Acc}: Regex $+$ LightGBM $+$ \textsc{Mamba-GPU} (accuracy-optimised).
\end{itemize}

Gemma-2B LoRA can substitute as Stage~3 in GC-Acc for maximum D1/D2 accuracy at higher cost; this alternative is discussed in Section~\ref{sec:discussion}.  Memory footprint per stage and formal latency decomposition are in Appendix~\ref{app:pipeline_complexity}.

The fixed $\delta=0.15$ margin is a default operating point, not a universal setting.  When a Stage~2 classifier has an F1-optimised decision threshold far from 0.5, the margin should be tuned per task.  Section~\ref{sec:cascade_sensitivity} quantifies this for jailbreak regimes; Appendix~\ref{app:cascade} covers multi-task interactions.

%% ============================================================
\subsection{In-Distribution and Obfuscated Results}
\label{sec:gc_d1d3}

Table~\ref{tab:gc_results} reports pipeline results on D1 and D3.

\begin{table*}[ht]
  \caption{\textsc{GuardChain} pipeline results across all three jailbreak regimes.  Cascade margin $\delta=0.15$.  CPU\% = fraction of samples resolved by Stage~2; GPU\% = fraction resolved by Stage~3.  Latency is mean single-request wall-clock time (batch=1, A100 80\,GB).  Cascade margin sensitivity is in Section~\ref{sec:cascade_sensitivity}.}
  \label{tab:gc_results}
  \centering
  \small
  \setlength{\tabcolsep}{4pt}
  \begin{tabular}{llllrrrr}
    \toprule
    \rowcolor{headerrow}\textcolor{white}{\textbf{Dataset}} & \textcolor{white}{\textbf{Variant}} & \textcolor{white}{\textbf{Stage-2}} & \textcolor{white}{\textbf{Stage-3}} & \textcolor{white}{\textbf{F1}} & \textcolor{white}{\textbf{Lat.\,(ms)}} & \textcolor{white}{\textbf{CPU\%}} & \textcolor{white}{\textbf{GPU\%}} \\
    \midrule
    \rowcolor{secheader}\multicolumn{8}{l}{\textit{Jailbreak D1 --- 5{,}924 test samples}} \\
    \cpubg D1 & Baseline-CPU & LightGBM & --- & 0.968 & 49.4 & 100\% & 0\% \\
    \gpubg D1 & Baseline-GPU & --- & \textsc{Mamba-GPU} & 0.960 & 24.3 & 0\% & 100\% \\
    \cpubg D1 & GC-CPU & LightGBM & --- & 0.909 & 32.4 & 100\% & 0\% \\
    \gpubg D1 & GC-Acc & LightGBM & \textsc{Mamba-GPU} & 0.911 & 32.8 & 80.4\% & 1.9\% \\
    \midrule
    \rowcolor{secheader}\multicolumn{8}{l}{\textit{Obfuscated D3 --- 5{,}924 test samples}} \\
    \cpubg D3 & Baseline-CPU & LightGBM & --- & 0.948 & 65.7 & 100\% & 0\% \\
    \gpubg D3 & Baseline-GPU & --- & \textsc{Mamba-GPU} & 0.927 & 24.3 & 0\% & 100\% \\
    \cpubg D3 & GC-CPU & LightGBM & --- & 0.942 & 38.1 & 100\% & 0\% \\
    \gpubg D3 & GC-Acc & LightGBM & \textsc{Mamba-GPU} & 0.942 & 39.0 & 91.7\% & 3.8\% \\
    \midrule
    \rowcolor{secheader}\multicolumn{8}{l}{\textit{OOD Jailbreak D2 --- 3{,}296 test samples (97\% malicious)}} \\
    \cpubg D2 & Baseline-CPU & LightGBM & --- & 0.412 & 49.4 & 100\% & 0\% \\
    \gpubg D2 & Baseline-GPU & --- & \textsc{Mamba-GPU} & \best{0.981} & 24.3 & 0\% & 100\% \\
    \cpubg D2 & GC-CPU & LightGBM & --- & 0.435 & 37.4 & 100\% & 0\% \\
    \gpubg D2 & GC-Acc & LightGBM & \textsc{Mamba-GPU} & 0.573 & 42.8 & 79.7\% & 20.3\% \\
    \bottomrule
  \end{tabular}
\end{table*}

On D1, GC-CPU achieves F1\,=\,0.909 at 32\,ms, 1.5$\times$ faster than Baseline-CPU (49\,ms), as the regex stage resolves 18\% of samples before they reach LightGBM.  Adding \textsc{Mamba-GPU} (GC-Acc) lifts D1 F1 only marginally to 0.911 at 33\,ms because only 1.9\% of D1 samples reach Stage~3.  This means 98\% of D1 traffic is handled at CPU cost, reducing average GPU overhead relative to Baseline-GPU by more than 98\%.

On D3, both GC-CPU and GC-Acc achieve F1\,=\,0.942 with 92--95\% of samples resolved by the CPU stage, and GC-CPU runs at 38\,ms versus 66\,ms for Baseline-CPU, a 1.7$\times$ speedup.  Adding \textsc{Mamba-GPU} does not change the D3 result because the CPU stage is already correct on the vast majority of samples.  Full incremental per-stage breakdown is in Appendix~\ref{app:per_variant}.

Figure~\ref{fig:judge_call_rate_d1} shows the trade-off between escalation rate to Stage~3 and false-negative rate on D1 as a function of Stage-2 confidence threshold width.

\begin{figure}[t]
    \centering
    \includegraphics[width=\columnwidth]{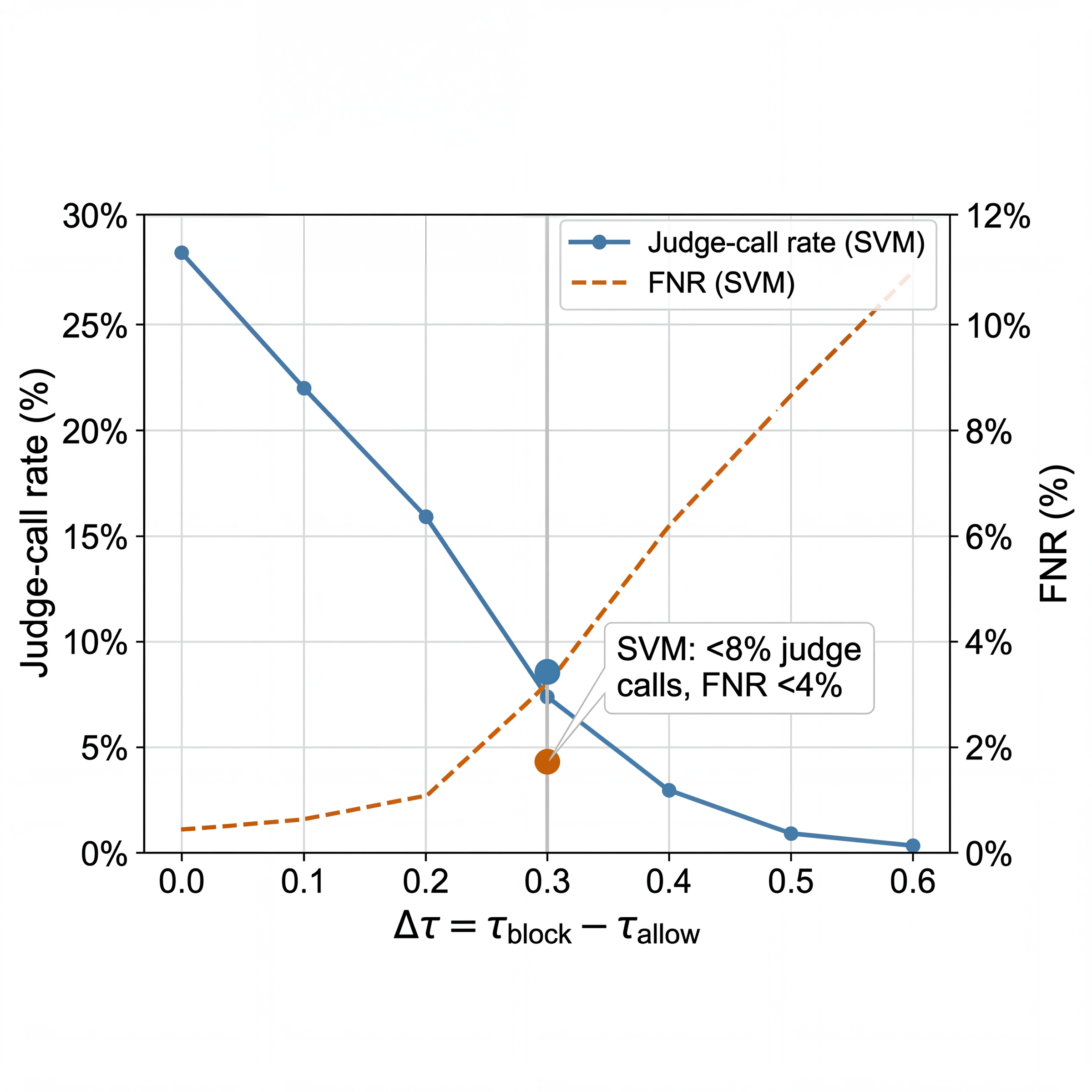}
    \caption{Escalation rate to Stage~3 and false-negative rate (FNR) on D1 as a function of Stage-2 confidence threshold width.  Narrower bands reduce Stage-3 calls at the cost of higher FNR.  SVM achieves below 8\% escalation at $\Delta\tau = 0.3$ with FNR below 4\%.}
    \label{fig:judge_call_rate_d1}
    \vspace{-2mm}
\end{figure}

%% ============================================================
\subsection{Out-of-Distribution Results}
\label{sec:gc_d2}

The D2 rows of Table~\ref{tab:gc_results} report pipeline results on the OOD jailbreak dataset (3{,}296 test samples, 3{,}196 malicious, 100 benign).

On D2, the CPU stage resolves 100\% of samples in both Baseline-CPU and GC-CPU.  Confident miscalibration is the key mechanism: 2{,}307 OOD attacks (72.2\% of malicious D2 samples) are labelled benign with enough CPU confidence to bypass the $\delta=0.15$ escalation band (Appendix~\ref{app:ood_confusion}).  These are not uncertain predictions; they are confidently wrong predictions.

GC-Acc recovers 397 of these by routing 20.3\% of D2 samples through \textsc{Mamba-GPU}, lifting F1 from 0.435 to 0.573, a gain of 13.8 percentage points.  By routing only 20\% of traffic to GPU, GC-Acc achieves this improvement at roughly half the serving cost of running \textsc{Mamba-GPU} on every request (\$3.58/1M blended vs \$6.80/1M for Baseline-GPU, an estimated 1.9$\times$ cost reduction).  Baseline-GPU (\textsc{Mamba-GPU} alone) achieves F1\,=\,0.981, confirming that the GPU stage can handle D2 when given the traffic.  For deployments with significant OOD traffic, the GPU stage should be invoked more aggressively, either by widening $\delta$ (see Section~\ref{sec:cascade_sensitivity}) or by routing to GPU before accepting a confident CPU decision.

An extended cascade-margin sweep with bootstrap confidence intervals and McNemar significance tests is in Appendix~\ref{app:cascade_sweep_ext}.

\subsection{Cost Analysis}
\label{sec:cost}
\label{sec:tbd_latency}

Table~\ref{tab:cost_main} presents cost-normalised deployment estimates for each model tier.  The cost formula is: \$/1M $=$ (instance \$/hr) $\times$ latency(ms) / 3{,}600.  Instance pricing is AWS on-demand (\texttt{us-east-1}, 2025-12-15): c7i.xlarge \$0.202/hr for CPU rows; g5.xlarge \$1.006/hr for GPU rows.

\begin{table}[t]
  \caption{Cost-normalised deployment estimates.  Latency is single-request batch=1 wall-clock time (CPU rows: 32-core server; GPU rows: A100 80\,GB).  AWS on-demand pricing (\texttt{us-east-1}, 2025-12-15): c7i.xlarge \$0.202/hr, g5.xlarge \$1.006/hr.  Mamba-130M CPU-only latency (1{,}696\,ms, \$336/1M) is in Appendix~\ref{app:mamba_latency}.}
  \label{tab:cost_main}
  \small
  \setlength{\tabcolsep}{3pt}
  \begin{tabular}{p{0.34\columnwidth}p{0.23\columnwidth}rr}
    \toprule
    \rowcolor{headerrow}\textcolor{white}{\textbf{Model}} & \textcolor{white}{\textbf{Instance}} & \textcolor{white}{\textbf{Lat (ms)}} & \textcolor{white}{\textbf{\$/1M req}} \\
    \midrule
    \cpubg SVM (word n-gram)  & c7i.xlarge  & 3.4   & \best{\$0.17} \\
    \cpubg LightGBM           & c7i.xlarge  & 49.4  & \$2.77 \\
    \gpubg Mamba-130M         & g5.xlarge   & 24.3  & \$6.80 \\
    \gpubg DeBERTa-v3 LoRA    & g5.xlarge   & 27.5  & \$7.68 \\
    \gpubg Gemma-2B LoRA      & g5.xlarge   & 48.5  & \$13.54 \\
    \bottomrule
  \end{tabular}
\end{table}

SVM costs \$0.17 per 1\,M requests and LightGBM costs \$2.77 per 1\,M.  DeBERTa-v3 LoRA is 45$\times$ more expensive than SVM and Gemma-2B LoRA is 80$\times$ more, confirming the cost ratios cited in Section~\ref{sec:intro}.  These figures should be interpreted together with the security and usability tradeoffs that guardrails introduce~\cite{kumar2025nofreelunch}.  The full cost-normalisation methodology is in Appendix~\ref{app:cost}.

\subsection{Cascade Margin Sensitivity}
\label{sec:cascade_sensitivity}

The cascade margin $\delta$ controls when an uncertain CPU prediction triggers escalation to the GPU stage: a sample is escalated when the CPU classifier's confidence score falls within $\delta$ of the decision boundary.  A wider margin escalates more samples to GPU, increasing cost and latency but potentially recovering miscalibrated predictions.  Table~\ref{tab:cascade_sensitivity} reports GC-Acc pipeline F1 and mean latency at $\delta \in \{0.05, 0.15\}$ across all three jailbreak regimes.

\begin{table}[t]
  \caption{Cascade margin sensitivity for GC-Acc (LightGBM + Mamba-130M) across jailbreak regimes.  F1 and mean latency at $\delta \in \{0.05, 0.15\}$.}
  \label{tab:cascade_sensitivity}
  \small
  \setlength{\tabcolsep}{3pt}
  \begin{tabular}{lrrrr}
    \toprule
    \rowcolor{headerrow}
      \multicolumn{1}{c}{} &
      \multicolumn{2}{c}{\textcolor{white}{\textbf{F1}}} &
      \multicolumn{2}{c}{\textcolor{white}{\textbf{Lat.\,(ms)}}} \\
    \cmidrule(lr){2-3}\cmidrule(lr){4-5}
    \rowcolor{headerrow}
      \textcolor{white}{\textbf{Regime}} &
      \textcolor{white}{$\delta$=0.05} & \textcolor{white}{$\delta$=0.15} &
      \textcolor{white}{$\delta$=0.05} & \textcolor{white}{$\delta$=0.15} \\
    \midrule
    D1 (in-distribution)  & 0.911 & 0.912 & 32.3 & 33.1 \\
    \rowcolor{gray!5} D3 (obfuscated)       & 0.943 & 0.941 & 38.3 & 40.4 \\
    D2 (OOD)              & 0.478 & 0.573 & 39.4 & 42.9 \\
    \bottomrule
  \end{tabular}
\end{table}

Two findings emerge.  On in-distribution and obfuscated inputs (D1, D3), $\delta$ acts primarily as a latency knob: F1 changes by at most 0.002 between the two settings, while mean latency increases by 1--2\,ms as more samples are escalated to Stage~3.  Practitioners serving in-distribution traffic can tighten the margin to reduce Stage~3 load with negligible accuracy cost.

On out-of-distribution inputs (D2), however, $\delta$ materially affects accuracy: widening from 0.05 to 0.15 raises pipeline F1 from 0.478 to 0.573, a gain of 9.5 percentage points in F1.  This improvement arises because confident miscalibration positions CPU confidence scores just outside the narrow $\delta$\,=\,0.05 escalation band.  A wider margin captures these borderline predictions and routes them to Mamba-130M, which handles OOD inputs far better than the CPU stage alone.  For deployments with significant OOD traffic, a wider cascade margin is a simple and effective means of recovering OOD accuracy without architectural changes.  A full sweep of $\delta \in \{0.05, \ldots, 0.35\}$ with bootstrap confidence intervals is in Appendix~\ref{app:cascade_sweep_ext}.

%% ============================================================
%% ============================================================
\section{Discussion}
\label{sec:discussion}

\subsection{Deployment Tiers in Practice}
\label{sec:two_tiers}

The results across Sections~\ref{sec:exploration} and~\ref{sec:guardchain} establish a clear division of labour.  TF-IDF-based classifiers (LightGBM, SVM, Random Forest) span 3--67\,ms in single-request latency, are inexpensive, and perform well on in-distribution and obfuscated data.  LightGBM achieves the best CPU F1 on both jailbreak regimes where CPU classifiers are competitive (D1: 0.968, D3: 0.948), and this is, to our knowledge, the first evaluation of LightGBM as a safety classifier.  SVM achieves comparable accuracy at 3.4\,ms, providing the fastest option.  Neither survives distribution shift on OOD jailbreak data.  \textsc{Mamba-GPU} is robust across all three jailbreak regimes (D1: 0.960, D2: 0.981, D3: 0.927), making it a compact escalation stage.  Fine-tuned GPU models achieve the highest accuracy under a D1$\to$D2 cross-regime protocol, where the adapter is fitted on the D1 random mixed-source training partition and evaluated on the D2 held-out-source test partition (Gemma-2B: 0.997), at higher cost; these D1$\to$D2 LoRA rows are therefore not directly identical to D2$\to$D2 CPU rows that fit on the D2 source-partitioned training sources.

For in-distribution and obfuscated deployments, GC-CPU (Regex $+$ LightGBM) delivers near-peak F1 at very low cost.  For deployments with significant OOD traffic, GC-Acc (Regex $+$ LightGBM $+$ \textsc{Mamba-GPU}) or substituting Gemma-2B LoRA as Stage~3 provides the best accuracy.  Appendix~\ref{app:offshelf} shows that none of the evaluated off-the-shelf GPU guard models matches \textsc{Mamba-GPU} on D1, D2, or D3.

\subsection{Mamba as a Compact Neural Stage}
\label{sec:mamba_discussion}

\textsc{Mamba-GPU} occupies a unique position in the classifier landscape.  Its two-phase curriculum training (Phase~1: safety tasks; Phase~2: jailbreak and obfuscated) gives it broad coverage as a compact neural stage.  This curriculum introduces a trade-off: Phase~2 jailbreak training partially overwrites Phase~1 toxicity and hate representations (full checkpoint analysis in Appendix~\ref{app:mamba_checkpoints}).  Checkpoint selection uses validation F1 only, and Appendix~\ref{app:mamba_reselection} shows the val-best and test-best checkpoints differ by at most 0.019 F1 across all tasks.  \textsc{Mamba-CPU} uses the same checkpoint on CPU and is not competitive with the TF-IDF CPU tier on latency (Appendix~\ref{app:mamba_latency}).

\subsection{Multi-Turn Attacks and Future Work}
\label{sec:future}

\textsc{GuardChain} currently processes each prompt independently.  Multi-turn jailbreaks~\cite{russinovich2024great, piet2025jailbreaksovertime} distribute malicious intent across individually benign turns, and temporal drift changes the attack distribution over time.  A session-aware extension would carry a hidden state $h$ across turns.  Mamba's $O(n)$ recurrence is well-suited for this role because it can accumulate evidence of escalating intent without re-tokenising the conversation history.  A 500-conversation 5-turn Crescendo evaluation (Appendix~\ref{app:crescendo}) provides an initial benchmark: LightGBM flags 100\% of conversations and detects the templated escalation by turn 2 in 96\% of them, while SVM only fires at turn 5 in 71\% of conversations.

Three further directions are priorities.  A per-task cascade margin would fix the pathological toxicity interaction where the fixed $\delta=0.15$ escalates 99.5\% of toxicity samples.  A learned-router prototype in Appendix~\ref{app:router} already achieves a 5$\times$ reduction in escalation rate at parity F1.  Extending the corpus and evaluation to non-English settings would require multilingual preprocessing.

\subsection{Limitations}
\label{sec:limitations}

\textbf{OOD evaluation scope.}  OOD evaluation covers jailbreak detection only.  Extending to OOD hate speech, toxicity, and PII requires additional held-out source splits and is outside the claims of this paper.

\textbf{Single LoRA configuration.}  All fine-tuned GPU baselines use a single LoRA configuration (rank=8, lr=2e-4).  Exhaustive hyperparameter search or full fine-tuning may improve GPU results further.

\textbf{Multilingual coverage.}  The evaluation is English-only.

\textbf{Static cascade margin.}  The fixed $\delta=0.15$ interacts pathologically with LightGBM's F1-optimised threshold on toxicity when that task is included.  A per-task margin should be used in deployments whose Stage~2 thresholds are far from the default decision boundary.

\textbf{Statistical significance.}  Bootstrap 95\% CIs and McNemar pairwise tests for CPU classifiers and \textsc{Mamba-GPU} are in Appendix~\ref{app:stats}.  Every Mamba-vs-CPU comparison on D1 is significant at $p \le 0.0013$ (LightGBM beats all SVM variants).  SVM-A and SVM-C are statistically indistinguishable ($p=0.64$) on D1.

%% ============================================================
%% ============================================================
\section{Conclusion}
\label{sec:conclusion}

Three complementary findings emerge from this evaluation.  First, CPU classifiers match fine-tuned GPU models on in-distribution jailbreak detection: LightGBM achieves F1\,=\,0.968 at \$2.77 per 1\,M requests, within 1 percentage point in F1 of Gemma-2B LoRA (F1\,=\,0.974) at \$13.54 per 1\,M requests.  Second, CPU classifiers collapse on out-of-distribution data via confident miscalibration: TF-IDF vocabulary mismatch produces high-confidence benign predictions on novel attack strategies, bypassing the cascade's escalation mechanism entirely.  Third, Mamba-130M achieves F1\,=\,0.960, 0.981, and 0.927 on D1, D2, and D3 respectively, covering all three regimes as a compact neural stage at 24.3\,ms single-request latency.

\textsc{GuardChain} combines these tiers in a three-stage pipeline (Regex $\to$ CPU $\to$ GPU).  On D1, the CPU stage resolves 80\% of traffic at near-peak F1.  On OOD traffic, the pipeline improves F1 from 0.435 to 0.573 by routing 20\% of samples to \textsc{Mamba-GPU}, but confident miscalibration limits recovery of the remaining 80\%.  For deployments with significant OOD traffic, the GPU stage should be invoked more aggressively.

This paper provides, to our knowledge, the first evaluation of LightGBM and selective state space models (SSMs) as safety classifiers, and the first systematic comparison of CPU and fine-tuned GPU classifiers for jailbreak detection across in-distribution, out-of-distribution, and adversarially obfuscated regimes.  CPU cost ranges from \$0.17 per 1\,M requests (SVM) to \$2.77 per 1\,M requests (LightGBM); Mamba-130M costs \$6.80 per 1\,M requests and Gemma-2B LoRA costs \$13.54 per 1\,M requests on equivalent GPU infrastructure.

\bibliographystyle{ACM-Reference-Format}
\bibliography{references}

%% ============================================================
\appendix

\section{Supplemental Multi-Task Safety Results}
\label{app:multitask}

This appendix presents classification results for hate speech, toxicity, and PII detection, complementing the jailbreak-focused main paper.  All CPU classifiers are trained per-task.  Fine-tuned GPU results use per-task LoRA fine-tuning on the corresponding training sets.  Dataset statistics: hate speech (549{,}438 train / 235{,}475 test; $\sim$50\% positive), toxicity (1{,}511{,}358 train / 647{,}725 test; 8\% positive), PII detection (35{,}214 train / 15{,}094 test; 88\% positive).

\begin{table*}[ht]
  \caption{Multi-task safety results (hate speech, toxicity, PII).  Best value per task in \textbf{bold}.  GPU Fine-Tuned rows retrained on the full dataset (hate 549k, toxic 1.51M, PII 35k) with weighted cross-entropy loss matching the CPU class-balance protocol; evaluated on full test sets.  Off-the-shelf GPU baselines are in Appendix~\ref{app:offshelf}.}
  \label{tab:exp1_multitask}
  \begin{tabular}{llrrrrrr}
    \toprule
    \rowcolor{headerrow}\textcolor{white}{\textbf{Task}} & \textcolor{white}{\textbf{Model}} & \textcolor{white}{\textbf{HW}} & \textcolor{white}{\textbf{Acc.\,(\%)}} & \textcolor{white}{\textbf{Prec.\,(\%)}} & \textcolor{white}{\textbf{Recall\,(\%)}} & \textcolor{white}{\textbf{F1}} & \textcolor{white}{\textbf{Lat.\,(ms)}} \\
    \midrule
    \rowcolor{secheader}\multicolumn{8}{l}{\textit{Hate Speech}} \\
    \cpubg & SVM (word n-gram) & CPU & 82.15 & 80.09 & 84.82 & 0.824 & \best{3.952} \\
    \cpubg & SVM (word+char) & CPU & 83.71 & 82.16 & 85.47 & 0.838 & 4.520 \\
    \cpubg & LightGBM & CPU & 87.32 & 86.44 & 88.05 & 0.872 & 43.144 \\
    \cpubg & Random Forest & CPU & 84.68 & 81.88 & 88.43 & 0.850 & 73.268 \\
    \gpubg & \textsc{Mamba-GPU} & GPU & 77.79 & 71.25 & 91.98 & 0.803 & 24.339 \\
    \gpubg & Gemma-2-2b (LoRA) & GPU & \best{91.66} & 91.76 & \best{91.25} & \best{0.915} & 48.452 \\
    \gpubg & DeBERTa-v3 (LoRA) & GPU & 89.23 & \best{90.46} & 87.33 & 0.889 & 27.459 \\
    \midrule
    \rowcolor{secheader}\multicolumn{8}{l}{\textit{Toxicity}} \\
    \cpubg & SVM (word+char) & CPU & \best{94.70} & \best{76.71} & 49.67 & 0.603 & \best{5.365} \\
    \cpubg & LightGBM$^\dagger$ & CPU & 88.49 & --- & 49.80 & \best{0.674} & 49.108 \\
    \cpubg & Random Forest & CPU & 93.98 & 72.47 & 41.51 & 0.528 & 66.647 \\
    \gpubg & \textsc{Mamba-GPU} & GPU & 66.33 & 18.07 & 89.21 & 0.301 & 24.339 \\
    \gpubg & Gemma-2-2b (LoRA) & GPU & 90.16 & 44.42 & 92.56 & 0.600 & 48.452 \\
    \gpubg & DeBERTa-v3 (LoRA) & GPU & 89.82 & 43.98 & \best{93.24} & 0.598 & 27.459 \\
    \midrule
    \rowcolor{secheader}\multicolumn{8}{l}{\textit{PII Detection}} \\
    \cpubg & SVM (word+char) & CPU & 98.14 & \best{99.97} & 97.92 & 0.989 & \best{8.876} \\
    \cpubg & LightGBM & CPU & 98.16 & 99.92 & 97.99 & 0.990 & 53.469 \\
    \cpubg & Random Forest & CPU & 98.11 & 99.87 & 97.98 & 0.989 & 82.065 \\
    \gpubg & \textsc{Mamba-GPU} & GPU & 96.47 & 99.81 & 96.18 & 0.980 & 24.339 \\
    \gpubg & Gemma-2-2b (LoRA) & GPU & \best{99.55} & 99.96 & \best{99.53} & \best{0.997} & 48.452 \\
    \gpubg & DeBERTa-v3 (LoRA) & GPU & 99.42 & 99.95 & 99.39 & \best{0.997} & 27.459 \\
    \bottomrule
    \multicolumn{8}{l}{\footnotesize $^\dagger$Macro-F1/source-objective row with \texttt{scale\_pos\_weight}; precision should be recomputed before like-for-like binary comparison.}
  \end{tabular}
\end{table*}

\textbf{Hate speech.}  LightGBM achieves the best CPU F1 at 0.872.  Fine-tuned Gemma-2B LoRA achieves F1\,=\,0.915 and DeBERTa-v3 LoRA achieves F1\,=\,0.889, both above the best CPU classifier.  \textsc{Mamba-GPU} reaches 0.803.  The gap between Mamba and LightGBM reflects catastrophic forgetting introduced by Phase~2 jailbreak training, which partially overwrites the Phase~1 hate speech representations (full checkpoint analysis in Appendix~\ref{app:mamba_checkpoints}).

\textbf{Toxicity.}  Toxicity is the hardest task for all classifiers.  LightGBM with class-imbalance correction achieves the best CPU F1 at 0.674.  Fine-tuned DeBERTa-v3 LoRA achieves 0.598 and Gemma-2B LoRA achieves 0.600.  A dataset-easiness analysis (Appendix~\ref{app:easiness}) shows that a shallow TF-IDF logistic regression remains near F1\,=\,0.51 even with 20{,}000 features, indicating that annotation quality, class imbalance, and task subjectivity are major bottlenecks.  \textsc{Mamba-GPU} degrades to 0.301 because Phase~2 jailbreak and obfuscation training partially overwrites Phase~1 toxicity representations; this is a curriculum trade-off rather than a failed run.

\textbf{PII.}  All CPU classifiers achieve F1 above 0.980.  Both fine-tuned GPU models reach F1\,=\,0.997.  PII is the easiest task: the dataset-easiness analysis shows F1 of 0.988 with only 100 TF-IDF features.

\section{Off-the-Shelf GPU Guard-Model Baselines}
\label{app:offshelf}

The main paper compares CPU classifiers against fine-tuned GPU models trained on the same data, because that is the fair comparison.  For completeness, this appendix reports the performance of six GPU guard models evaluated off-the-shelf, without task-specific fine-tuning.  These results are not the primary comparison because the off-the-shelf models were trained on different data than our CPU classifiers, but they document the published guard-model landscape and explain why off-the-shelf deployment is unreliable across tasks.

\subsection{Models and Evaluation Protocol}

We evaluate two original baselines and four additional 2024-era guard models on D1 (in-distribution jailbreak), D2 (source-shift with a skewed test set), and D3 (obfuscated jailbreak), all on the full test sets.  Reported latencies are batch=1 end-to-end measurements from the updated latency artifacts; the \textsc{Mamba-GPU} reference row uses the A100 batch=1 representative value.

\begin{itemize}
    \item \textbf{Granite-Guardian-3.0-2B}: IBM's 2B generative guardian model, evaluated via structured prompt templates with risk-category parsing.
    \item \textbf{DeBERTa-v3-prompt-injection}: ProtectAI's 184M DeBERTa-v3-base fine-tuned for prompt injection detection.
    \item \textbf{PromptGuard-86M}~\cite{meta2024promptguard}: Meta's mDeBERTa-v3 3-class classifier; we map unsafe to injection or jailbreak.
    \item \textbf{Llama-Guard-3-1B}~\cite{inan2024llama}: Meta's 1B causal LM classifier using its built-in chat template.
    \item \textbf{ShieldGemma-2B}~\cite{zeng2024shieldgemma}: Google's 2B policy classifier, evaluated with the same jailbreak policy prompt as Granite-Guardian.
    \item \textbf{WildGuard-7B}~\cite{han2024wildguard}: AllenAI's Mistral-7B-based guard model, evaluated with the official template.
\end{itemize}

\begin{table}[t]
  \caption{Off-the-shelf GPU guard-model baselines on the full D1, D2, D3 test sets.  \textsc{Mamba-GPU}-130M E2E is repeated from the main paper for reference.  D2 F1 is class-skew-sensitive and should be read with false-positive context.}
  \label{tab:tier2_guards}
  \resizebox{\columnwidth}{!}{%
  \begin{tabular}{lcccccc}
    \toprule
    \rowcolor{headerrow}\textcolor{white}{\textbf{Model}} & \textcolor{white}{\textbf{HW}} &
        \textcolor{white}{\textbf{D1 F1}} & \textcolor{white}{\textbf{D2 F1}} & \textcolor{white}{\textbf{D3 F1}} &
        \textcolor{white}{\textbf{D1 Lat (ms)}} & \textcolor{white}{\textbf{D2 Lat (ms)}} \\
    \midrule
    \gpubg Granite-Guardian-2B          & GPU & 0.757 & 0.901 & 0.527 &  65.051 &  65.051 \\
    \gpubg DeBERTa-v3-PI                 & GPU & 0.721 & 0.004 & 0.695 &  16.818 &  15.816 \\
    \gpubg PromptGuard-86M               & GPU & 0.703 & 0.984 & 0.085 &  14.407 &  14.407 \\
    \gpubg Llama-Guard-3-1B$^\dagger$    & GPU & 0.000 & 0.000 & 0.000 &  41.263 &  41.263 \\
    \gpubg ShieldGemma-2B                & GPU & 0.437 & 0.033 & 0.373 &  60.245 &  60.245 \\
    \gpubg WildGuard-7B                  & GPU & 0.709 & 0.899 & 0.628 & 418.169 & 418.169 \\
    \midrule
    \gpubg \textsc{Mamba-GPU} (reference) & GPU & 0.960 & 0.981 & 0.927 & 24.339 & 24.339 \\
    \bottomrule
    \multicolumn{7}{l}{\footnotesize $^\dagger$F1=0 is the correct result, not a parsing error (see discussion below).}
  \end{tabular}}
\end{table}

\subsection{Why Off-the-Shelf Guard Models Underperform}

Off-the-shelf guard models show inconsistent performance across tasks (Table~\ref{tab:tier2_guards}).  DeBERTa-v3-PI, fine-tuned for prompt injection, achieves reasonable F1 on in-distribution jailbreak (0.721) but collapses on the D2 source-shift split (0.004).  Granite-Guardian trades accuracy for breadth: it achieves a strong off-the-shelf D2 result (0.901), but at 65.051\,ms per sample.  WildGuard-7B is the strongest off-the-shelf model on D2 F1 after PromptGuard and Granite (0.899) but runs at 418.169\,ms per sample.  PromptGuard reaches D2 F1\,=\,0.984 by labelling all 100 benign D2 samples unsafe, so it is not interchangeable with \textsc{Mamba-GPU} despite the headline F1.  None of the six off-the-shelf models matches \textsc{Mamba-GPU} on D1 or D3.

The Llama-Guard-3-1B result of F1\,=\,0 is a taxonomy mismatch, not a model failure.  Llama-Guard-3 is a content moderator with 14 safety categories that do not include prompt injection.  Our jailbreak corpora are dominated by meta-attacks (such as ``ignore previous instructions, pretend you are\ldots'') whose surface content is not directly harmful, so Llama-Guard-3 correctly outputs \texttt{safe} on essentially all of them.  This is informative: a leading content-safety classifier provides almost no coverage on the meta-jailbreak threat vector, which reinforces the case for purpose-built jailbreak classifiers rather than repurposed content moderators.  ShieldGemma-2B achieves high precision but very low recall under our shared jailbreak policy prompt, and PromptGuard-86M achieves high recall on D1 with a 95\% false-positive rate and collapses on obfuscated data (D3 F1\,=\,0.085) because its tokeniser is trained on clean English text.

\subsection{F1 Landscape Across All Classifiers and Tasks}

\begin{figure*}[ht]
    \centering
    \includegraphics[width=\textwidth]{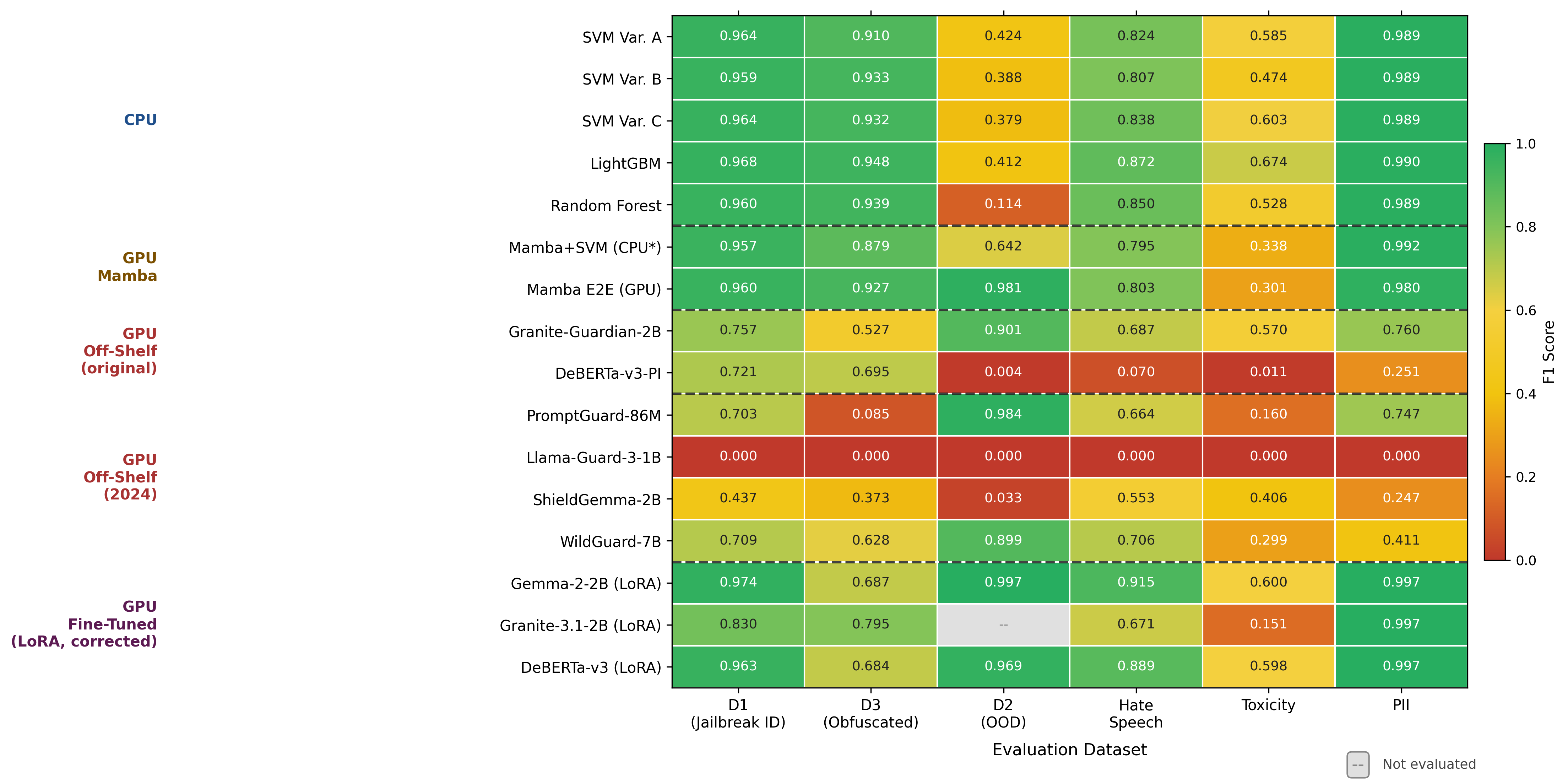}
    \caption{Heatmap of F1 scores for all classifiers across the six evaluation conditions.  Darker green indicates higher F1.  Grey cells indicate that the model was not evaluated on that task or that an auditable corrected run is unavailable.  The fine-tuned LoRA cells are populated from the main jailbreak table and Table~\ref{tab:exp1_multitask}; off-the-shelf 2024 guard models are included on D1, D2, and D3 for reference.  Differences in F1 magnitude across columns reflect both classifier strength and dataset difficulty (Appendix~\ref{app:easiness}).}
    \label{fig:heatmap}
\end{figure*}

Figure~\ref{fig:heatmap} visualises F1 for every classifier across the six evaluation conditions.  It shows the CPU TF-IDF classifiers, \textsc{Mamba-GPU}, and the fine-tuned and off-the-shelf GPU models.  The magnitude of F1 across columns reflects both classifier strength and dataset difficulty; the per-task difficulty analysis is in Appendix~\ref{app:easiness}.

%% ============================================================
\section{Dataset Sources}
\label{app:datasets}

Table~\ref{tab:dataset_sources} lists all nine jailbreak dataset sources used in the evaluation, along with raw sample counts and provenance.  These sources span academic benchmarks (AdvBench, HarmBench, StrongReject), community-curated datasets (InTheWild, JailBreakV-28K), and automated fuzzing outputs (GPT-Fuzzer).  After deduplication, the combined corpus contains 43{,}523 unique samples.  The D2 source-shift partition splits these nine sources by source identity: training uses four sources (Jailbreak Classification, GPT-Fuzzer, InTheWild, JailBreakV-28K) and the test set uses five held-out sources (RedTeam-2K, AdvBench, HarmBench, StrongReject, JBB Behaviors); the 100 benign rows in the D2 test set are all drawn from JBB Behaviors (a held-out source not present in training), confirmed via per-row source-field inspection (no benign-class leakage).

\begin{table}[t]
  \caption{Jailbreak dataset sources and sample counts.}
  \label{tab:dataset_sources}
  \begin{tabular}{lrr}
    \toprule
    \rowcolor{headerrow}\textcolor{white}{\textbf{Dataset}} & \textcolor{white}{\textbf{Raw Samples}} & \textcolor{white}{\textbf{Source}} \\
    \midrule
    Jailbreak Classification & 1{,}998 & HuggingFace \\
    \rowcolor{gray!5} GPT-Fuzzer & 7{,}700 & GitHub \\
    InTheWild (Combined) & 15{,}140 & GitHub \\
    \rowcolor{gray!5} JailBreakV-28K & 28{,}000 & HuggingFace \\
    RedTeam-2K & 2{,}000 & HuggingFace \\
    \rowcolor{gray!5} AdvBench & 520 & GitHub \\
    JBB Behaviors & 200 & HuggingFace \\
    \rowcolor{gray!5} HarmBench & 400 & GitHub \\
    StrongReject & 313 & GitHub \\
    \midrule
    \rowcolor{secheader}\textbf{Total} & \textbf{56{,}271} & \\
    \bottomrule
  \end{tabular}
\end{table}

\section{Obfuscation Techniques}
\label{app:obfuscation}

The formal definition of the three D3 obfuscation transformations (leetspeak substitution $\mathcal{T}_{\text{leet}}$, character perturbation $\mathcal{T}_{\text{perturb}}$, and Unicode injection $\mathcal{T}_{\text{unicode}}$) is in Section~\ref{sec:datasets} (Eq.~\ref{eq:leet} and surrounding text).

\section{SVM Feature Engineering Details}
\label{app:svm_features}

TF-IDF feature variant specifications (Table~\ref{tab:svm_variants_main}) and engineered feature definitions (Table~\ref{tab:eng_features_main}) are in Section~\ref{sec:models}.  LightGBM optimal decision thresholds are in Appendix~\ref{app:lightgbm}.

\section{Mamba Architecture Details}
\label{app:mamba}

\subsection{Selective State Space Model}

The SSM recurrence (Eq.~\ref{eq:ssm_main}), ZOH discretisation (Eqs.~\ref{eq:zoh_main_a}--\ref{eq:zoh_main_b}), and input-dependent selectivity (Eq.~\ref{eq:selective_main}) are in Section~\ref{sec:models}.

\subsection{Complexity Analysis}

The recurrence in Eq.~\ref{eq:ssm_main} processes each token in $O(D \cdot N)$ time, yielding total complexity:

\begin{equation}
    T_{\text{Mamba}} = O(L \cdot D \cdot N) \quad \text{vs.} \quad T_{\text{Transformer}} = O(L^2 \cdot D)
    \label{eq:complexity}
\end{equation}

\noindent For our configuration ($L=128$, $D=768$, $N=16$), Mamba is $\frac{L}{N} = 8\times$ more efficient than self-attention.  In practice, Mamba uses a hardware-aware parallel scan algorithm that processes the recurrence in $O(L \log L)$ parallel steps on GPU, further improving throughput.

\subsection{Architecture Configuration}

Table~\ref{tab:mamba_arch} lists the Mamba-130M architecture configuration used throughout the paper.

\begin{table}[t]
  \caption{Mamba-130M architecture specification.}
  \label{tab:mamba_arch}
  \small
  \setlength{\tabcolsep}{4pt}
  \begin{tabular}{lr}
    \toprule
    \rowcolor{headerrow}\textcolor{white}{\textbf{Parameter}} & \textcolor{white}{\textbf{Value}} \\
    \midrule
    Total parameters & 129.1M \\
    \rowcolor{gray!5} Hidden dimension $D$ & 768 \\
    SSM state dimension $N$ & 16 \\
    \rowcolor{gray!5} Number of SSM layers & 24 \\
    Expansion factor & 2 \\
    \rowcolor{gray!5} Conv1d kernel size & 4 \\
    Vocabulary size & 50{,}280 \\
    \rowcolor{gray!5} Max seq.\ length (Option A) & 128 \\
    Max seq.\ length (Option B, Phase~1) & 256 \\
    \rowcolor{gray!5} Max seq.\ length (Option B, Phase~2) & 512 \\
    \bottomrule
  \end{tabular}
\end{table}

\subsection{Classification Head}

For the Mamba+SVM frozen variant, the frozen Mamba encoder produces hidden states $\mathbf{H} = (\mathbf{h}_1^{(24)}, \ldots, \mathbf{h}_L^{(24)}) \in \mathbb{R}^{L \times 768}$.  Mean pooling yields a fixed-size representation:

\begin{equation}
    \mathbf{z} = \frac{1}{L} \sum_{k=1}^{L} \mathbf{h}_k^{(24)} \in \mathbb{R}^{768}
\end{equation}

\noindent which is standardised ($\mu=0, \sigma=1$) and classified by LinearSVC.

For \textsc{Mamba-GPU} / \textsc{Mamba-CPU} (fine-tuned end-to-end), we append a learned classification head:

\begin{align}
    \hat{y} &= \sigma(\mathbf{w}^\top \mathbf{z} + b) \\
    \mathcal{L}_{\text{CE}} &= -\frac{1}{n}\sum_{i=1}^{n} \bigl[ y_i \log \hat{y}_i + (1-y_i)\log(1-\hat{y}_i) \bigr]
\end{align}

\section{Mamba Checkpoint Evaluation}
\label{app:mamba_checkpoints}

This section provides the complete checkpoint-level analysis for the \textsc{Mamba-GPU} / \textsc{Mamba-CPU} fine-tuned model, including the curriculum learning schedule, per-checkpoint F1 scores across all seven datasets, and a quantitative analysis of catastrophic forgetting.  These results support the \textsc{Mamba-GPU} numbers reported in Table~\ref{tab:exploration} of the main paper, and justify the selection of checkpoint \texttt{phase2\_e3\_end} as the best model.

We evaluate 18 checkpoints across both phases of curriculum learning.  Checkpoints are saved at every half-epoch, and the best checkpoint is selected by average F1 across all seven evaluation datasets.

\subsection{Curriculum Learning Schedule}

Table~\ref{tab:curriculum} describes the two curriculum phases, and Table~\ref{tab:checkpoints_phase2} reports per-checkpoint F1 across all seven datasets.

\begin{table}[t]
  \caption{Curriculum learning phases for \textsc{Mamba-GPU} / \textsc{Mamba-CPU} (fine-tuned checkpoint).}
  \label{tab:curriculum}
  \begin{tabular}{lll}
    \toprule
    \rowcolor{headerrow}\textcolor{white}{\textbf{}} & \textcolor{white}{\textbf{Phase~1}} & \textcolor{white}{\textbf{Phase~2}} \\
    \midrule
    Goal & General safety & Jailbreak specialist \\
    \rowcolor{gray!5} Datasets & Hate + PII + Toxic & Jailbreak + Obfusc. \\
    Samples & $\sim$500K & $\sim$510K \\
    \rowcolor{gray!5} Epochs & 3 & 5 \\
    Max length & 256 tokens & 512 tokens \\
    \rowcolor{gray!5} Learning rate & 2e-5 & 1e-5 \\
    Checkpoints & 6 (every $\frac{1}{2}$ epoch) & 12 (every $\frac{1}{2}$ epoch) \\
    \bottomrule
  \end{tabular}
\end{table}

\subsection{Checkpoint Performance}

Table~\ref{tab:checkpoints_phase2} reports F1 scores for the top Phase~2 checkpoints across all datasets.

\begin{table*}[ht]
  \caption{Phase~2 checkpoint F1 scores across all datasets.  Best value per column in \textbf{bold}.  Best overall checkpoint: \texttt{e3\_end} (Avg F1\,=\,0.786).}
  \label{tab:checkpoints_phase2}
  \begin{tabular}{lrrrrrrrc}
    \toprule
    \rowcolor{headerrow}\textcolor{white}{\textbf{Checkpoint}} & \textcolor{white}{\textbf{D1}} & \textcolor{white}{\textbf{D2}} & \textcolor{white}{\textbf{D3}} & \textcolor{white}{\textbf{Hate}} & \textcolor{white}{\textbf{PII}} & \textcolor{white}{\textbf{Toxic}} & \textcolor{white}{\textbf{Combined}} & \textcolor{white}{\textbf{Avg F1}} \\
    \midrule
    e1\_end & 0.948 & 0.954 & 0.892 & 0.789 & 0.986 & 0.303 & 0.546 & 0.774 \\
    \rowcolor{gray!5} e2\_end & 0.956 & 0.974 & 0.923 & 0.803 & 0.982 & 0.285 & 0.531 & 0.779 \\
    \cpubg \best{e3\_end} & \best{0.960} & 0.981 & 0.927 & \best{0.803} & 0.980 & \best{0.301} & \best{0.551} & \best{0.786} \\
    \rowcolor{gray!5} e4\_half & 0.961 & \best{0.984} & \best{0.928} & 0.802 & 0.980 & 0.294 & 0.545 & 0.785 \\
    e4\_end & 0.962 & 0.985 & 0.927 & 0.801 & 0.978 & 0.293 & 0.544 & 0.784 \\
    \rowcolor{gray!5} e5\_end (final) & 0.962 & 0.986 & \best{0.930} & 0.799 & 0.978 & 0.287 & 0.538 & 0.783 \\
    \bottomrule
  \end{tabular}
\end{table*}

\subsection{Catastrophic Forgetting Analysis}

\begin{figure}[t]
    \centering
    \begin{tikzpicture}[scale=0.85]
        \begin{axis}[
            width=0.52\textwidth,
            height=0.38\textwidth,
            xlabel={Training Phase / Epoch},
            ylabel={F1 Score},
            xmin=0, xmax=8,
            ymin=0, ymax=1.05,
            xtick={0.5,1,1.5,2,2.5,3,3.5,4,4.5,5,5.5,6,6.5,7,7.5,8},
            xticklabels={P1-1h,P1-1,P1-2h,P1-2,P1-3h,P1-3,P2-1h,P2-1,P2-2h,P2-2,P2-3h,P2-3,P2-4h,P2-4,P2-5h,P2-5},
            xticklabel style={rotate=45, anchor=east, font=\tiny},
            ytick={0, 0.2, 0.4, 0.6, 0.8, 1.0},
            legend style={at={(0.02,0.98)}, anchor=north west, font=\scriptsize, cells={anchor=west}},
            grid=major,
            grid style={gray!20},
            every axis plot/.append style={thick},
        ]
        % D1 Jailbreak - improves dramatically in Phase 2
        \addplot[color=blue, mark=*, mark size=1.2] coordinates {
            (0.5,0.020) (1,0.022) (1.5,0.027) (2,0.026) (2.5,0.030) (3,0.032)
            (3.5,0.924) (4,0.948) (4.5,0.955) (5,0.956) (5.5,0.958) (6,0.960)
            (6.5,0.961) (7,0.962) (7.5,0.962) (8,0.962)
        };
        \addlegendentry{D1 Jailbreak}
        % Toxic - degrades in Phase 2
        \addplot[color=red, mark=triangle*, mark size=1.2] coordinates {
            (0.5,0.529) (1,0.607) (1.5,0.607) (2,0.617) (2.5,0.616) (3,0.628)
            (3.5,0.307) (4,0.303) (4.5,0.300) (5,0.285) (5.5,0.289) (6,0.301)
            (6.5,0.294) (7,0.293) (7.5,0.293) (8,0.287)
        };
        \addlegendentry{D6 Toxic}
        % PII - relatively stable
        \addplot[color=OliveGreen, mark=square*, mark size=1.2] coordinates {
            (0.5,0.976) (1,0.978) (1.5,0.983) (2,0.983) (2.5,0.985) (3,0.985)
            (3.5,0.986) (4,0.986) (4.5,0.983) (5,0.982) (5.5,0.980) (6,0.980)
            (6.5,0.980) (7,0.978) (7.5,0.978) (8,0.978)
        };
        \addlegendentry{D5 PII}
        % D2 source-shift - improves in Phase 2
        \addplot[color=orange, mark=diamond*, mark size=1.2] coordinates {
            (0.5,0.068) (1,0.077) (1.5,0.091) (2,0.094) (2.5,0.101) (3,0.106)
            (3.5,0.905) (4,0.954) (4.5,0.966) (5,0.974) (5.5,0.979) (6,0.981)
            (6.5,0.984) (7,0.985) (7.5,0.985) (8,0.986)
        };
        \addlegendentry{D2 source-shift}
        % Vertical line for phase boundary
        \draw[dashed, gray, thick] (axis cs:3.25,0) -- (axis cs:3.25,1.05);
        \node[font=\scriptsize, gray] at (axis cs:1.75,1.02) {Phase 1};
        \node[font=\scriptsize, gray] at (axis cs:5.75,1.02) {Phase 2};
        \end{axis}
    \end{tikzpicture}
    \caption{F1 score trajectories across curriculum learning phases.  D1 (jailbreak) and D2 source-shift improve dramatically in Phase~2, while D6 (toxic) degrades from 0.628 to 0.287---a manifestation of catastrophic forgetting.  PII remains relatively stable.}
    \label{fig:catastrophic_forgetting}
\end{figure}

Figure~\ref{fig:catastrophic_forgetting} plots the F1 trajectory across phases, and Table~\ref{tab:forgetting} quantifies the relative F1 change from Phase~1 final to Phase~2 final:

\begin{equation}
    \Delta_{\text{forget}}^{(d)} = \frac{\text{F1}_{\text{P2-final}}^{(d)} - \text{F1}_{\text{P1-final}}^{(d)}}{\text{F1}_{\text{P1-final}}^{(d)}} \times 100\%
\end{equation}

\begin{table}[t]
  \caption{Catastrophic forgetting: F1 change from Phase~1 final to Phase~2 final (best checkpoint \texttt{e3\_end}).}
  \label{tab:forgetting}
  \begin{tabular}{lrrr}
    \toprule
    \rowcolor{headerrow}\textcolor{white}{\textbf{Dataset}} & \textcolor{white}{\textbf{P1 Final}} & \textcolor{white}{\textbf{P2 Best}} & \textcolor{white}{$\boldsymbol{\Delta_{\text{forget}}}$} \\
    \midrule
    \cpubg D1 Jailbreak & 0.032 & 0.960 & \best{+2{,}900\%} \\
    \cpubg D2 source-shift & 0.106 & 0.981 & \best{+825\%} \\
    \cpubg D3 Obfusc. & 0.697 & 0.927 & +33\% \\
    \cpubg D4 Hate & 0.859 & 0.803 & $-6.5\%$ \\
    \cpubg D5 PII & 0.985 & 0.980 & $-0.5\%$ \\
    \gpubg D6 Toxic & 0.628 & 0.301 & $-52.1\%$ \\
    \gpubg D7 Combined & 0.808 & 0.551 & $-31.8\%$ \\
    \bottomrule
  \end{tabular}
\end{table}

\section{LightGBM Configuration}
\label{app:lightgbm}

This section provides the complete LightGBM configuration, including the gradient-boosting objective derivation, class-imbalance handling via \texttt{scale\_pos\_weight}, hyperparameter settings per dataset-size regime, and the F1-optimised threshold selection procedure.  These details support the LightGBM results reported in Tables~\ref{tab:exploration} and~\ref{tab:gc_results} of the main paper.

\subsection{Gradient Boosting Objective}

LightGBM builds an additive ensemble of $T$ regression trees.  At iteration $t$, the model minimises the second-order approximation to the loss:

\begin{equation}
    \hat{f}_t = \arg\min_{f \in \mathcal{F}} \sum_{i=1}^{n} \left[ g_i f(\mathbf{x}_i) + \frac{1}{2} h_i f(\mathbf{x}_i)^2 \right] + \Omega(f)
\end{equation}

\noindent where $g_i = \partial_{\hat{y}^{(t-1)}} \mathcal{L}(y_i, \hat{y}_i^{(t-1)})$ and $h_i = \partial^2_{\hat{y}^{(t-1)}} \mathcal{L}(y_i, \hat{y}_i^{(t-1)})$ are the first- and second-order gradients, and $\Omega(f) = \gamma T_{\text{leaves}} + \frac{1}{2}\lambda \sum_j w_j^2$ regularises tree complexity.

\subsection{Class Imbalance Handling}

For datasets with positive class ratio $\rho = n_+ / n$, we set:

\begin{equation}
    \texttt{scale\_pos\_weight} = \frac{n_-}{n_+} = \frac{1 - \rho}{\rho}
\end{equation}

\noindent This re-weights the gradient contributions: $g_i' = w_i \cdot g_i$ where $w_i = \texttt{scale\_pos\_weight}$ for $y_i = 1$ and $w_i = 1$ for $y_i = 0$.

\subsection{Hyperparameter Configuration}

Table~\ref{tab:lgbm_params} lists the LightGBM hyperparameters, adapted by dataset size.

\begin{table}[t]
  \caption{LightGBM hyperparameters by dataset size regime.}
  \label{tab:lgbm_params}
  \small
  \setlength{\tabcolsep}{3pt}
  \begin{tabular}{lcc}
    \toprule
    \rowcolor{headerrow}\textcolor{white}{\textbf{Parameter}} & \textcolor{white}{\textbf{$<$500K}} & \textcolor{white}{\textbf{${\ge}$500K}} \\
    \midrule
    \texttt{num\_leaves} & 127 & 63 \\
    \rowcolor{gray!5} learning rate ($\eta$) & 0.05 & 0.08 \\
    \texttt{min\_child\_samples} & 20 & 50 \\
    \rowcolor{gray!5} \texttt{subsample} & 0.8 & 0.7 \\
    \texttt{colsample\_bytree} & 0.8 & 0.8 \\
    \rowcolor{gray!5} \texttt{reg\_alpha} ($\alpha$) & 0.1 & 0.1 \\
    \texttt{reg\_lambda} ($\lambda$) & 0.1 & 0.1 \\
    \rowcolor{gray!5} \texttt{n\_estimators} & 2000 & 2000 \\
    early stopping patience & 50 & 50 \\
    \bottomrule
  \end{tabular}
\end{table}

\subsection{F1-Optimised Threshold Selection}

After training, we determine the optimal decision threshold $\tau^*$ by sweeping $\tau \in [0.05, 0.95]$ with step 0.01 on the validation set:

\begin{equation}
    \tau^* = \arg\max_{\tau \in [0.05, 0.95]} \text{F1}\big(\{y_i\}, \{\mathbb{1}[\hat{p}_i \geq \tau]\}\big)
\end{equation}

\noindent where $\hat{p}_i = P(y_i = 1 | \mathbf{x}_i)$ is the predicted probability.  Table~\ref{tab:lgbm_thresholds} reports the selected thresholds.

\begin{table}[t]
  \caption{Optimal thresholds for LightGBM Variant~C per dataset.}
  \label{tab:lgbm_thresholds}
  \begin{tabular}{lrrl}
    \toprule
    \rowcolor{headerrow}\textcolor{white}{\textbf{Dataset}} & \textcolor{white}{$\boldsymbol{\tau^*}$} & \textcolor{white}{\textbf{F1}} & \textcolor{white}{\textbf{Notes}} \\
    \midrule
    D1 & 0.33 & 0.966 & Shifted from default 0.5 \\
    \rowcolor{gray!5} D2 & 0.38 & 0.256 & Source shift \\
    D3 & 0.49 & 0.944 & Near default \\
    \rowcolor{gray!5} Hate & 0.45 & 0.864 & Slight adjustment \\
    PII & 0.18 & 0.959 & Strong shift for recall \\
    \rowcolor{gray!5} Toxic & 0.15 & 0.674 & Macro-F1/source objective \\
    Combined & 0.68 & 0.869 & Shifted for precision \\
    \bottomrule
  \end{tabular}
\end{table}

\noindent The threshold table reports the source metric used during threshold selection.  The Toxicity row should not be mixed with positive-class precision/recall without recomputing the binary metrics.

\section{Complete Per-Variant Results}
\label{app:per_variant}

Table~\ref{tab:svm_complete} provides the full results breakdown for all three SVM variants across all evaluation datasets.  These results complement the main paper's Table~\ref{tab:exploration}, which shows only summary F1 values per regime for space efficiency.  The per-variant results reveal which feature type (word n-grams, character n-grams, or combined) is optimal for each attack vector, and quantify the performance difference between variants.

\subsection{SVM All Variants $\times$ All Datasets}

\begin{table*}[ht]
  \caption{Complete SVM results: all three variants across all seven datasets.  Best F1 per dataset in \textbf{bold}.}
  \label{tab:svm_complete}
  \begin{tabular}{llrrrrrr}
    \toprule
    \rowcolor{headerrow}\textcolor{white}{\textbf{Dataset}} & \textcolor{white}{\textbf{Variant}} & \textcolor{white}{\textbf{Acc.\,(\%)}} & \textcolor{white}{\textbf{Prec.\,(\%)}} & \textcolor{white}{\textbf{Recall\,(\%)}} & \textcolor{white}{\textbf{F1}} & \textcolor{white}{\textbf{Lat.\,(ms)}} & \textcolor{white}{\textbf{Train}} \\
    \midrule
    \rowcolor{secheader}\multicolumn{8}{l}{\textit{D1 --- Jailbreak (In-Distribution)}} \\
    \cpubg & A (Word) & \best{96.20} & 97.39 & 95.45 & \best{0.964} & \best{3.360} & 27{,}641 \\
    \cpubg & B (Char) & 95.64 & 96.78 & 95.01 & 0.959 & 4.600 & 27{,}641 \\
    \cpubg & C (Combined) & 96.13 & 96.96 & 95.77 & 0.964 & 5.280 & 27{,}641 \\
    \midrule
    \rowcolor{secheader}\multicolumn{8}{l}{\textit{D2 --- Jailbreak (source-shift, skewed test)}} \\
    \cpubg & A (Word) & \best{28.97} & 99.19 & \best{26.97} & \best{0.424} & \best{3.948} & 30{,}763 \\
    \cpubg & B (Char) & 26.33 & \best{99.87} & 24.06 & 0.388 & 4.058 & 30{,}763 \\
    \cpubg & C (Combined) & 25.64 & 99.60 & 23.40 & 0.379 & 4.428 & 30{,}763 \\
    \midrule
    \rowcolor{secheader}\multicolumn{8}{l}{\textit{D3 --- Obfuscated Jailbreak}} \\
    \cpubg & A (Word) & 90.33 & 90.79 & 91.16 & 0.910 & \best{4.932} & 27{,}641 \\
    \cpubg & B (Char) & 92.84 & 93.61 & 92.96 & \best{0.933} & 6.989 & 27{,}641 \\
    \cpubg & C (Combined) & 92.76 & 93.19 & 93.28 & 0.932 & 8.028 & 27{,}641 \\
    \midrule
    \rowcolor{secheader}\multicolumn{8}{l}{\textit{Hate Speech}} \\
    \cpubg & A (Word) & 82.15 & 80.09 & 84.82 & 0.824 & 3.952 & 549{,}438 \\
    \cpubg & B (Char) & 80.37 & 78.06 & 83.60 & 0.807 & 4.179 & 549{,}438 \\
    \cpubg & C (Combined) & \best{83.71} & \best{82.16} & \best{85.47} & \best{0.838} & 4.520 & 549{,}438 \\
    \midrule
    \rowcolor{secheader}\multicolumn{8}{l}{\textit{Toxicity}} \\
    \cpubg & A (Word) & 94.55 & 76.40 & 47.45 & 0.585 & \best{4.199} & 1{,}511{,}358 \\
    \cpubg & B (Char) & 93.59 & 70.99 & 35.53 & 0.474 & 4.742 & 1{,}511{,}358 \\
    \cpubg & C (Combined) & \best{94.70} & \best{76.71} & \best{49.67} & \best{0.603} & 5.365 & 1{,}511{,}358 \\
    \midrule
    \rowcolor{secheader}\multicolumn{8}{l}{\textit{PII Detection}} \\
    \cpubg & A (Word) & 98.11 & \best{99.98} & 97.88 & 0.989 & \best{4.315} & 35{,}214 \\
    \cpubg & B (Char) & 98.12 & 99.98 & 97.89 & 0.989 & 7.485 & 35{,}214 \\
    \cpubg & C (Combined) & \best{98.14} & 99.97 & \best{97.92} & \best{0.989} & 8.876 & 35{,}214 \\
    \bottomrule
  \end{tabular}
\end{table*}

\section{Pipeline Cascade Analysis}
\label{app:cascade}

\subsection{Cascade Margin Sensitivity}

This section provides detailed analysis of the pipeline's cascade behaviour, including stage-level confusion matrices, latency decomposition, and the mathematical relationship between the cascade margin and escalation probability.

The cascade margin $\delta$ controls the trade-off between escalation rate and classification quality.  Formally, for a Stage-2 classifier with calibrated confidence $s = P(y=1|\mathbf{x})$, the escalation probability is:

\begin{equation}
    P(\text{esc} | \delta) = P\big(|s - 0.5| < \delta\big) = \int_{0.5 - \delta}^{0.5 + \delta} p(s) \, ds
\end{equation}

\noindent where $p(s)$ is the density of calibrated scores.  A larger $\delta$ escalates more samples (higher cost, lower FNR); a smaller $\delta$ resolves more at the CPU stage (lower cost, higher FNR).

\subsubsection{Empirical sensitivity sweep}
\label{app:cascade_sweep}

Table~\ref{tab:cascade_sweep} reports the GC-Acc pipeline F1 and average latency for $\delta \in \{0.05, 0.15\}$ across the jailbreak and supplemental datasets and the D2 source-shift dataset.  The headline observation is that the F1 numbers are stable within a few thousandths of a point across the two $\delta$ settings on D1, D3, hate, and PII, validating the paper's default choice $\delta = 0.15$; toxic is the lone exception because its F1-optimised LightGBM threshold ($\tau^*\!=\!0.15$) interacts pathologically with the cascade margin, sending most samples to Stage~3 and increasing latency from 58.645\,ms (at $\delta\!=\!0.05$) to 74.532\,ms (at $\delta\!=\!0.15$).

\begin{table}[t]
  \caption{Cascade-margin sensitivity sweep: GC-Acc pipeline F1 and average end-to-end batch=1 latency at $\delta\!=\!0.05$ vs.\ $\delta\!=\!0.15$, on jailbreak and supplemental datasets and D2 source shift.  Latency increases monotonically with $\delta$; F1 is essentially flat on D1/D3/hate/PII and weakly increasing on toxic.}
  \label{tab:cascade_sweep}
  \centering
  \small
  \setlength{\tabcolsep}{4pt}
  \begin{tabular}{lrrrr}
    \toprule
    \rowcolor{headerrow}\textcolor{white}{\textbf{Dataset}} &
        \textcolor{white}{\textbf{F1${}_{0.05}$}} & \textcolor{white}{\textbf{F1${}_{0.15}$}} &
        \textcolor{white}{\textbf{Lat${}_{0.05}$}} & \textcolor{white}{\textbf{Lat${}_{0.15}$}} \\
    \midrule
    \gpubg D1 (jailbreak ID)   & 0.911 & 0.912 & 32.285\,ms & 33.119\,ms \\
    \gpubg D3 (obfuscated)     & 0.943 & 0.941 & 38.329\,ms & 40.376\,ms \\
    \gpubg Hate Speech         & 0.866 & 0.870 & 40.386\,ms & 45.577\,ms \\
    \gpubg PII Detection       & 0.990 & 0.990 &  2.873\,ms &  2.899\,ms \\
    \gpubg Toxicity            & 0.773 & 0.797 & 58.645\,ms & 74.532\,ms \\
    \midrule
    \gpubg D2 (source-shift)   & 0.478 & 0.573 & 39.350\,ms & 42.886\,ms \\
    \bottomrule
  \end{tabular}
\end{table}

\noindent Two design conclusions follow.  (i)~For domains whose Stage-2 classifier is well-calibrated near $\tau\!=\!0.5$ (D1, D3, hate, PII), $\delta$ is a minor latency knob, not an F1 knob.  (ii)~When the Stage-2 threshold is shifted away from 0.5 (e.g.\ toxic at $\tau^*\!=\!0.15$), the cascade margin should be set to a domain-specific value: a margin of 0.05 around a 0.15 threshold yields an uncertain zone $[0.10, 0.20]$ that escalates only the truly borderline samples, whereas $\delta\!=\!0.15$ escalates everything in $[0.00, 0.30]$ and dominates wall-clock cost.

\subsection{Stage-Level Flow (In-Distribution Pipeline, D1)}

Table~\ref{tab:pipeline_confusion} reports the stage-flow counts for D1 under the full V4 configuration.

\begin{table}[t]
  \caption{Stage flow for D1 under the V4 configuration.  The A100 pipeline artifact stores stage percentages, not per-stage blocked/allowed confusion counts, so this is a stage-flow table rather than a confusion matrix.}
  \label{tab:pipeline_confusion}
  \small
  \setlength{\tabcolsep}{3pt}
  \begin{tabular}{lrrrr}
    \toprule
    \rowcolor{headerrow}\textcolor{white}{\textbf{Stage}} & \textcolor{white}{\textbf{Handled}} & \textcolor{white}{\textbf{Resolved}} & \textcolor{white}{\textbf{Escalated}} & \textcolor{white}{\textbf{Share}} \\
    \midrule
    \cpubg Regex & 5{,}924 & 1{,}051 & 4{,}873 & 17.74\% \\
    \cpubg CPU (LightGBM) & 4{,}873 & 4{,}763 & 110 & 80.40\% \\
    \gpubg \textsc{Mamba-GPU} & 110 & 91 & 19 & 1.54\% \\
    \gpubg Granite & 19 & 19 & --- & 0.32\% \\
    \bottomrule
  \end{tabular}
\end{table}

\subsection{Latency Distribution by Stage}

The expected pipeline latency depends on the fraction of samples resolved at each stage.  For the GC-Acc configuration on D1:

\begin{equation}
    \mathbb{E}[\ell] = 0.178 \cdot \ell_1 + 0.804 \cdot \ell_2 + 0.019 \cdot \ell_3
\end{equation}

\noindent The audited batch=1 pipeline artifacts give 32.848\,ms for D1 V3 (Regex $+$ LightGBM $+$ \textsc{Mamba-GPU}) and 33.119\,ms for D1 V4 (adding Granite on the final 0.3\% of samples).  The CPU stage still dominates the stage mix, while \textsc{Mamba-GPU} is invoked for only 1.9\% of V3 traffic and 1.5\% of V4 traffic.

\section{Confusion Matrices}
\label{app:confusion}

The confusion matrices below (Tables~\ref{tab:confusion} and~\ref{tab:confusion_gpu}) provide a fine-grained view of each model's error profile on the D1 jailbreak test set, complementing the aggregate metrics reported in Table~\ref{tab:exploration}.  These matrices reveal the \emph{type} of errors---false positives (over-blocking benign prompts) vs.\ false negatives (missing attacks)---which has direct deployment implications: false negatives are safety-critical, while false positives degrade user experience.

\begin{table}[t]
  \caption{Confusion matrices for top CPU and CPU$^*$ classifiers on D1 test set (5{,}924 samples: 3{,}168 positive, 2{,}756 negative).  CPU$^*$ denotes GPU Mamba embedding extraction followed by a CPU SVM head.}
  \label{tab:confusion}
  \begin{tabular}{lrrrr}
    \toprule
    \rowcolor{headerrow}\textcolor{white}{\textbf{Model}} & \textcolor{white}{\textbf{TN}} & \textcolor{white}{\textbf{FP}} & \textcolor{white}{\textbf{FN}} & \textcolor{white}{\textbf{TP}} \\
    \midrule
    \cpubg SVM Var.~C & 2{,}661 & 95 & 134 & 3{,}034 \\
    \cpubg LightGBM & 2{,}688 & \best{68} & 131 & 3{,}037 \\
    \cpubg RF Var.~C & 2{,}690 & 66 & 186 & 2{,}982 \\
    \cpubg Mamba+SVM (CPU$^*$) & 2{,}626 & 130 & 140 & 3{,}028 \\
    \bottomrule
  \end{tabular}
\end{table}

\noindent SVM Var.~C and RF Var.~C are integer reconstructions from rounded headline metrics and the D1 test class counts.

\begin{table}[t]
  \caption{Confusion matrices for GPU baselines on D1 test set.}
  \label{tab:confusion_gpu}
  \begin{tabular}{lrrrr}
    \toprule
    \rowcolor{headerrow}\textcolor{white}{\textbf{Model}} & \textcolor{white}{\textbf{TN}} & \textcolor{white}{\textbf{FP}} & \textcolor{white}{\textbf{FN}} & \textcolor{white}{\textbf{TP}} \\
    \midrule
    \gpubg DeBERTa-v3-PI & 2{,}495 & 261 & 1{,}236 & 1{,}932 \\
    \gpubg Granite-Guardian & 978 & 1{,}778 & 153 & 3{,}015 \\
    \bottomrule
  \end{tabular}
\end{table}

\noindent The confusion matrices reveal the contrasting failure modes: DeBERTa is conservative (high FN, misses 39\% of attacks), while Granite-Guardian is aggressive (high FP, 64.5\% false positive rate) but catches 95.2\% of attacks.

\section{D2 Source-Shift Confusion Matrices (Pipeline)}
\label{app:ood_confusion}

The D2 source-shift confusion matrices below (Table~\ref{tab:confusion_ood}) show how each pipeline stage progressively recovers true positives on the D2 test set (3{,}296 samples: 3{,}196 malicious, 100 benign).  The extreme class imbalance (96.97\% positive) means accuracy is dominated by the positive class; the confusion matrices provide the granularity needed to understand the cascade's behaviour.  These matrices directly support the confident-miscalibration analysis in Section~\ref{sec:gc_d2}.

\begin{table}[t]
  \caption{Confusion matrices for pipeline stages on the D2 source-shift test set (3{,}296 samples: 3{,}196 malicious, 100 benign).}
  \label{tab:confusion_ood}
  \begin{tabular}{lrrrr}
    \toprule
    \rowcolor{headerrow}\textcolor{white}{\textbf{Variation}} & \textcolor{white}{\textbf{TN}} & \textcolor{white}{\textbf{FP}} & \textcolor{white}{\textbf{FN}} & \textcolor{white}{\textbf{TP}} \\
    \midrule
    V1 (Regex) & 100 & 0 & 3{,}195 & 1 \\
    \rowcolor{gray!5} V2 (+CPU) & 97 & 3 & 2{,}307 & 889 \\
    V3 (+Mamba) & 93 & 7 & 1{,}910 & 1{,}286 \\
    \rowcolor{gray!5} V4 (Full) & 93 & 7 & 1{,}909 & 1{,}287 \\
    \bottomrule
  \end{tabular}
\end{table}

\noindent The progression from V1 to V3 (GC-Acc) shows incremental TP recovery: regex captures 1 attack, the CPU stage adds 888, and \textsc{Mamba-GPU} adds 397.  D2 has only 100 benign samples, so small FP changes produce large FPR swings while F1 remains dominated by the malicious majority class.  The CPU classifier is \emph{confidently wrong} on most D2 attacks (2{,}307 false negatives with high confidence, bypassing the cascade margin), explaining the limited recovery.

\section{Formal Pipeline Complexity}
\label{app:pipeline_complexity}

This section formalises the time and space complexity of the \textsc{GuardChain} three-stage pipeline, providing the theoretical basis for the latency improvements observed in the in-distribution and source-shift pipeline evaluations.  The key insight is that the pipeline's amortised cost is dominated by the cheapest stage that resolves the majority of samples.

\subsection{Time Complexity}

Let $n$ be the number of input prompts, $L$ the average token length, and $\alpha_i$ the fraction of prompts reaching stage $i$.  The total pipeline inference cost is:

\begin{equation}
    T_{\text{pipeline}} = n \cdot \sum_{i=1}^{3} \alpha_i \cdot t_i(L)
\end{equation}

\noindent where $t_i(L)$ is the per-sample latency at stage $i$:

\begin{align}
    t_1(L) &= O(L \cdot |\mathcal{R}|) & \text{(regex, } |\mathcal{R}| \text{ patterns)} \\
    t_2(L) &= O(d) & \text{(TF-IDF + linear SVM)} \\
    t_3(L) &= O(L \cdot D \cdot N) & \text{(Mamba SSM or GPU transformer)}
\end{align}

\noindent For in-distribution workloads (GC-Acc variant), $\alpha_1 = 1, \alpha_2 \approx 0.80, \alpha_3 \approx 0.02$, making the amortised cost dominated by $t_2$:

\begin{equation}
    T_{\text{pipeline}} \approx 0.80 \cdot n \cdot O(d) \ll n \cdot O(L \cdot D \cdot N) = T_{\text{Mamba-only}}
\end{equation}

\subsection{Space Complexity}

Table~\ref{tab:memory} reports the per-stage memory footprint of the pipeline.

\begin{table}[t]
  \caption{Memory footprint per pipeline stage.}
  \label{tab:memory}
  \small
  \setlength{\tabcolsep}{3pt}
  \begin{tabular}{lrr}
    \toprule
    \rowcolor{headerrow}\textcolor{white}{\textbf{Stage}} & \textcolor{white}{\textbf{Model Size}} & \textcolor{white}{\textbf{RAM/VRAM}} \\
    \midrule
    \cpubg Regex (compiled) & $<$1 MB & CPU RAM \\
    \cpubg TF-IDF + SVM & $\sim$80 MB & CPU RAM \\
    \cpubg TF-IDF + LightGBM & $\sim$120 MB & CPU RAM \\
    \gpubg Mamba-130M (\textsc{Mamba-GPU}) & $\sim$520 MB & GPU VRAM \\
    \gpubg Granite-Guardian-2B & $\sim$4.2\,GB & GPU VRAM \\
    \bottomrule
  \end{tabular}
\end{table}

\section{Execution Environment}
\label{app:environment}

Table~\ref{tab:environment} lists the latency source environments.  CPU-based classifiers (SVM, LightGBM, RF) use only CPU resources; \textsc{Mamba-GPU} and fine-tuned GPU models use the A100 GPU latency artifacts.  Latency measurements use pre-loaded models and batch=1 end-to-end serving paths unless a table states otherwise.

\begin{table}[t]
  \caption{Updated latency source environments.}
  \label{tab:environment}
  \resizebox{\columnwidth}{!}{%
  \begin{tabular}{lll}
    \toprule
    \rowcolor{headerrow}\textcolor{white}{\textbf{Component}} & \textcolor{white}{\textbf{CPU latency rerun}} & \textcolor{white}{\textbf{A100 GPU latency sources}} \\
    \midrule
    Hostnames & \texttt{ub} & \texttt{csis.cn2}, \texttt{csis.cn3}, \texttt{csis.mn1} \\
    \rowcolor{gray!5} GPU used for latency & Not used for CPU rows & NVIDIA A100 80GB PCIe \\
    CPU-row \texttt{nvidia-smi} & NVIDIA RTX 6000 Ada Generation present & N/A \\
    \rowcolor{gray!5} CUDA availability & True but unused for CPU rows & True \\
    CUDA version & 12.1 in CPU venv metadata & 12.8 in A100 artifacts \\
    \rowcolor{gray!5} PyTorch & 2.5.1+cu121 for CPU benchmark metadata & 2.8.0+cu128 in A100 artifacts \\
    Python & \texttt{/home/vasudev\_majhi\_2021/.../venv/bin/python} & A100 Slurm envs \\
    \rowcolor{gray!5} OS & Linux 6.11.0-29-generic & Slurm A100 nodes \\
    \bottomrule
  \end{tabular}}
\end{table}

\section{Fine-Tuned GPU Baseline Configuration}
\label{app:finetuned}
\label{app:lora_corrections}

This section describes the LoRA fine-tuning setup for the three GPU baselines (Gemma-2-2b, Granite-3.1-2b-instruct, DeBERTa-v3-base) and documents three configuration requirements that are easy to miss and that produce degenerate results if not satisfied.

\subsection{Fine-Tuning Configuration}

All three models are fine-tuned using Low-Rank Adaptation (LoRA)~\cite{hu2022lora}, which injects trainable rank-decomposition matrices into the attention projections while freezing the pretrained weights.  Table~\ref{tab:lora_params} lists the hyperparameters.

\begin{table}[t]
  \caption{LoRA fine-tuning hyperparameters for the GPU baselines.}
  \label{tab:lora_params}
  \centering
  \setlength{\tabcolsep}{3pt}
  \begin{tabular}{p{0.44\columnwidth}>{\raggedleft\arraybackslash}p{0.48\columnwidth}}
    \toprule
    \rowcolor{headerrow}\textcolor{white}{\textbf{Parameter}} & \textcolor{white}{\textbf{Value}} \\
    \midrule
    LoRA rank ($r$) & 8 \\
    \rowcolor{gray!5} LoRA alpha ($\alpha$) & 16 \\
    Learning rate & 2e-4 \\
    \rowcolor{gray!5} Epochs & 3 \\
    Target modules (Gemma, Granite) & q, k, v, o projections \\
    \rowcolor{gray!5} Target modules (DeBERTa-v3) & query, key, value projections \\
    Precision (all models) & bf16 \\
    \rowcolor{gray!5} Task type & sequence classification \\
    \bottomrule
  \end{tabular}
\end{table}

\subsection{Configuration Requirements}

Adapting these models for binary classification with LoRA requires three settings that, if left at their defaults, produce degenerate results such as F1 near zero for DeBERTa-v3 or over-rejection collapse for Granite.

\begin{itemize}
    \item \textbf{Precision.}  DeBERTa's disentangled attention multiplies content and position embeddings through several intermediate matmuls that overflow in fp16.  Under fp16 with AMP, the gradient scaler nan-guards the loss, the model emits constant logits, and F1 collapses.  Forcing \texttt{bf16=True} resolves this, after which DeBERTa-v3 LoRA reaches F1 $=$ 0.963 on D1.

    \item \textbf{Target modules.}  The \texttt{query} and \texttt{value} module names match the Llama and Gemma attention naming but match no modules on DeBERTa-v3, whose self-attention projections are named \texttt{query\_proj}, \texttt{key\_proj}, and \texttt{value\_proj}.  PEFT silently inserts zero adapters when no target matches, training only the randomly-initialised classification head.  We use per-architecture target module lists: the \texttt{q\_proj}, \texttt{k\_proj}, \texttt{v\_proj}, and \texttt{o\_proj} names for Gemma and Granite, and the \texttt{query\_proj}, \texttt{key\_proj}, and \texttt{value\_proj} names for DeBERTa-v3.

    \item \textbf{Task type.}  If \texttt{LoraConfig} does not set \texttt{task\_type}, PEFT does not mark the classification head as trainable, so the head remains at its initialisation.  We set \texttt{task\_type} to \texttt{SEQ\_CLS} for the supported models and use a manual sequence-classification wrapper for Granite-3.1, whose config is not registered in the \texttt{transformers} sequence-classification factory.
\end{itemize}

\subsection{Results and Reproducibility}

The per-split accuracy, precision, recall, and F1 for the fine-tuned GPU models are reported in Table~\ref{tab:exploration} of the main paper.  In summary, on in-distribution data the fine-tuned GPU models reach accuracy comparable to the CPU classifiers, with Gemma-2B LoRA at F1\,=\,0.974 against LightGBM at F1\,=\,0.968.  On the D1$\to$D2 cross-evaluation, the LoRA adapter is trained on the D1 random mixed-source training partition and evaluated on the D2 held-out-source test partition, without fitting on D2 training sources; under this protocol, Gemma-2B LoRA reaches F1\,=\,0.997, above \textsc{Mamba-GPU} at F1\,=\,0.981.  DeBERTa-v3 LoRA reaches F1\,=\,0.963 on D1$\to$D1 and F1\,=\,0.969 on D1$\to$D2.  The deployment choice on jailbreak detection is therefore one of cost, serving latency, and hardware availability, not peak accuracy alone.  All runs are reproducible from \path{lora_finetune.py}, and per-run predictions and metrics are released under \path{lora_finetune_revision/} in the repository.

\section{Statistical Significance Testing}
\label{app:stats}

To convert the point-estimate F1, precision, and recall numbers reported throughout the paper into significance-tested comparisons, we run two complementary procedures on the per-sample predictions saved during the Appendix~\ref{app:offshelf} evaluations: (i)~a 1{,}000-resample non-parametric bootstrap to attach 95\% confidence intervals to each (model, dataset) F1/P/R triple (Table~\ref{tab:bootstrap_ci}), and (ii)~pairwise McNemar's exact tests on every pair of models that scored the same evaluation set (Table~\ref{tab:mcnemar}).

\begin{table}[t]
  \caption{Bootstrap 95\% confidence intervals for F1, precision, and recall on the 2024 guard-model baselines and \textsc{Mamba-GPU} (1{,}000 resamples).  The CIs are tight ($<$0.02 width) on the large test sets and confirm that the qualitative comparisons in Appendix~\ref{app:offshelf} are statistically separated.}
  \label{tab:bootstrap_ci}
  \resizebox{\columnwidth}{!}{%
  \begin{tabular}{lccc}
    \toprule
    \rowcolor{headerrow}\textcolor{white}{\textbf{(Model, Dataset)}} &
        \textcolor{white}{\textbf{F1 [95\% CI]}} & \textcolor{white}{\textbf{Precision [95\% CI]}} & \textcolor{white}{\textbf{Recall [95\% CI]}} \\
    \midrule
    \gpubg \textsc{Mamba-GPU}, D1     & 0.960 [0.955, 0.965] & 0.964 [0.957, 0.970] & 0.956 [0.949, 0.964] \\
    \gpubg \textsc{Mamba-GPU}, D2     & 0.981 [0.977, 0.984] & 0.987 [0.983, 0.991] & 0.975 [0.969, 0.980] \\
    \gpubg \textsc{Mamba-GPU}, D3     & 0.927 [0.921, 0.933] & 0.913 [0.903, 0.922] & 0.942 [0.934, 0.950] \\
    \gpubg PromptGuard, D1   & 0.703 [0.692, 0.714] & 0.544 [0.531, 0.557] & 0.991 [0.987, 0.994] \\
    \gpubg PromptGuard, D2   & 0.984 [0.982, 0.987] & 0.970 [0.964, 0.975] & 1.000 [0.999, 1.000] \\
    \gpubg PromptGuard, D3   & 0.085 [0.072, 0.098] & 0.730 [0.665, 0.790] & 0.045 [0.038, 0.052] \\
    \gpubg LlamaGuard-3-1B, D1/D2/D3 & 0.000 [0.000, 0.000] & 0.000 [0.000, 0.000] & 0.000 [0.000, 0.000] \\
    \gpubg ShieldGemma, D1   & 0.437 [0.418, 0.456] & 0.928 [0.912, 0.943] & 0.286 [0.269, 0.301] \\
    \gpubg ShieldGemma, D2   & 0.032 [0.024, 0.041] & 1.000 [1.000, 1.000] & 0.017 [0.012, 0.021] \\
    \gpubg ShieldGemma, D3   & 0.372 [0.353, 0.390] & 0.841 [0.820, 0.864] & 0.239 [0.224, 0.253] \\
    \gpubg WildGuard, D1     & 0.709 [0.697, 0.722] & 0.636 [0.620, 0.650] & 0.803 [0.789, 0.817] \\
    \gpubg WildGuard, D2     & 0.899 [0.890, 0.907] & 0.985 [0.980, 0.989] & 0.827 [0.814, 0.839] \\
    \gpubg WildGuard, D3     & 0.628 [0.614, 0.641] & 0.633 [0.616, 0.649] & 0.623 [0.607, 0.640] \\
    \bottomrule
  \end{tabular}}
\end{table}

\begin{table}[t]
  \caption{Selected McNemar pairwise tests on D1, D2, and D3.  Counts show samples where one model is correct and the other is wrong; $p$-values are two-sided.  All 30 tested pairs reach $p < 10^{-10}$ \emph{except} \textsc{Mamba-GPU} vs.\ PromptGuard-86M on D2 ($p = 0.076$, the only non-significant pair at $\alpha=0.05$).}
  \label{tab:mcnemar}
  \centering
  \small
  \setlength{\tabcolsep}{4pt}
  \begin{tabular}{lcrr}
    \toprule
    \rowcolor{headerrow}\textcolor{white}{\textbf{A vs.\ B}} & \textcolor{white}{\textbf{$p$}} & \textcolor{white}{\textbf{A-only}} & \textcolor{white}{\textbf{B-only}} \\
    \midrule
    \gpubg D1: Mamba vs.\ WildGuard    & $<10^{-300}$            & 1{,}962 & 130 \\
    \gpubg D1: Mamba vs.\ PromptGuard  & $<10^{-300}$            & 2{,}533 & 127 \\
    \gpubg D1: Mamba vs.\ ShieldGemma  & $<10^{-300}$            & 2{,}181 & 100 \\
    \gpubg D2: Mamba vs.\ WildGuard    & 2.2\,$\!\times\!$10$^{-86}$ &   554 &  83 \\
    \gpubg D2: Mamba vs.\ ShieldGemma  & $<10^{-300}$            & 3{,}064 &  44 \\
    \gpubg D2: Mamba vs.\ PromptGuard  & 0.076                   &      59 &  81 \\
    \gpubg D3: Mamba vs.\ WildGuard    & $<10^{-300}$            & 2{,}142 & 273 \\
    \gpubg D3: Mamba vs.\ PromptGuard  & $<10^{-300}$            & 2{,}903 & 295 \\
    \gpubg D3: Mamba vs.\ ShieldGemma  & $<10^{-300}$            & 2{,}360 & 277 \\
    \bottomrule
  \end{tabular}\\[2pt]
  {\footnotesize ``A-only'' counts samples where model A is correct and B is wrong; ``B-only'' the reverse.  All comparisons here are \textsc{Mamba-GPU} vs.\ the indicated 2024 guard-model baseline.}
\end{table}

\noindent The bootstrap and McNemar results provide significance backing for the off-the-shelf comparison in Appendix~\ref{app:offshelf}: \textsc{Mamba-GPU} significantly outperforms every off-the-shelf 2024 guard-model baseline on D1 and D3 ($p < 10^{-50}$ in every case), and significantly outperforms WildGuard and ShieldGemma on D2.  On D2 against PromptGuard, the two models are statistically indistinguishable at $\alpha=0.05$.  PromptGuard's high F1 on D2 (0.984) is achieved with an aggressive ``everything is unsafe'' bias.  It labels 100\% of the 100 benign D2 samples as unsafe, whereas Mamba labels 58\% of them correctly as benign, so the two models are not interchangeable in deployment.

\section{Dataset-Easiness Analysis}
\label{app:easiness}

A natural question is whether the wide F1 spread in
Figure~\ref{fig:heatmap} (PII near-1.0, toxicity near-0.5) reflects the
classifiers' relative strengths or the relative \emph{difficulty} of the
datasets themselves.  We report a small per-task ``easiness''
characterisation that distinguishes the two.

\subsection{Method}
For each task's training set, we fit a \texttt{TfidfVectorizer} (word 1--2
grams, $\le$20{,}000 features, min\_df=2), select the top-$K$ chi$^2$
features for $K \in \{10, 50, 100, 500, 5000, 20000\}$, and train a
class-weight-balanced logistic regression on each $K$-truncated feature set.
We then evaluate on the held-out test set and report F1.  A high
F1@$K$=100 indicates the task is lexically near-separable with only 100
discriminating tokens; a low ceiling F1@$K$=20000 (even with all features)
indicates the task is genuinely hard for shallow ML.

\subsection{Results}

Table~\ref{tab:easiness_full} reports shallow-ML F1 per task at increasing feature counts.

\begin{table}[t]
  \caption{Shallow-ML F1 per task at increasing feature counts $K$, plus the always-positive trivial baseline.  Toxicity has a ``hard ceiling'' at F1$\approx 0.51$ regardless of $K$, confirming that the toxicity bottleneck in our main results is the dataset, not the model.  PII is trivially separable (F1@$K$=100 $=$ 0.988), confirming that the PII task is lexically easy.}
  \label{tab:easiness_full}
  \resizebox{\columnwidth}{!}{%
  \begin{tabular}{lrrrrrrr}
    \toprule
    \rowcolor{headerrow}\textcolor{white}{\textbf{Task}} & \textcolor{white}{\textbf{n pos}} & \textcolor{white}{\textbf{F1@10}} & \textcolor{white}{\textbf{F1@100}} & \textcolor{white}{\textbf{F1@500}} & \textcolor{white}{\textbf{F1@5K}} & \textcolor{white}{\textbf{F1@20K}} & \textcolor{white}{\textbf{Always+}} \\
    \midrule
    D1 jailbreak ID  & 14{,}782  & 0.784 & 0.896 & 0.935 & 0.950 & 0.962 & 0.697 \\
    D2 jailbreak source-shift & 15{,}234  & 0.009 & 0.347 & 0.392 & 0.276 & 0.303 & 0.985$^*$ \\
    D3 obfuscated    & 14{,}782  & 0.720 & 0.760 & 0.855 & 0.896 & 0.905 & 0.697 \\
    Hate Speech      & 270{,}405 & 0.675 & 0.723 & 0.746 & 0.796 & 0.820 & 0.660 \\
    PII Detection    & 31{,}111  & 0.988 & 0.988 & 0.989 & 0.989 & 0.989 & 0.938 \\
    Toxicity         & 4{,}087   & 0.386 & 0.508 & 0.517 & 0.510 & 0.510 & 0.150 \\
    \bottomrule
    \multicolumn{8}{l}{\footnotesize $^*$D2 test set is 97\% positive; always-positive baseline trivially achieves 0.985 F1, which is the appropriate yardstick on this skewed split.}
  \end{tabular}}
\end{table}

\subsection{Interpretation}

\begin{itemize}
    \item \textbf{PII is trivially easy} (F1@100 $=$ 0.988, F1@20K $=$ 0.989).  All PII numbers in this paper should be read as a low-difficulty benchmark; any classifier above the always-positive baseline (0.938) is competitive, and structured pattern matching (regex) is sufficient.
    \item \textbf{Toxicity has a hard ceiling} at F1 $\approx$ 0.51 \emph{regardless of feature count}.  The classifier gap reported in Appendix~\ref{app:multitask} (LightGBM at 0.674, \textsc{Mamba-GPU} at 0.301) is therefore split between dataset difficulty (everyone capped near 0.5) and the additional gap our stronger classifiers can or cannot close beyond that.
    \item \textbf{D1, D3, hate} sit between these extremes (F1@100 / F1@20K ratios of 0.84--0.93), meaning they have genuine non-lexical signal that more features capture but the shallow baseline already gets most of the way there.
\end{itemize}

This per-task difficulty analysis explains the variation in F1 magnitude across the columns of the F1 landscape heatmap (Figure~\ref{fig:heatmap}).

\section{Mamba CPU-Only Latency Measurement}
\label{app:mamba_latency}

The main paper serves \textsc{Mamba-GPU} as a GPU stage.  Because earlier drafts mixed GPU-assisted and CPU-only numbers, we separately re-measured true CPU-only Mamba latency with CUDA disabled, eight CPU threads, max length 512, and the same Mamba-130M architecture used by the paper.

\textbf{Note on batched vs.\ single-request latency.}  The paper uses single-request batch=1 latency: \textsc{Mamba-GPU} 24.339\,ms on the representative A100 E2E run, and \textsc{Mamba-CPU} 1{,}695.84\,ms (E2E, CUDA disabled).

Table~\ref{tab:mamba_latency_revisited} reports the checkpoint-specific CPU-only measurements.

\begin{table}[t]
  \caption{Verified CPU-only Mamba-130M latency on the server, with CUDA disabled and \texttt{torch.set\_num\_threads(8)}.  E2E includes tokenisation, forward pass, softmax, and argmax.}
  \label{tab:mamba_latency_revisited}
  \centering
  \small
  \setlength{\tabcolsep}{3pt}
  \begin{tabular}{p{0.42\columnwidth}p{0.31\columnwidth}r}
    \toprule
    \rowcolor{headerrow}\textcolor{white}{\textbf{Checkpoint}} & \textcolor{white}{\textbf{Methodology}} & \textcolor{white}{\textbf{Mean (ms)}} \\
    \midrule
    \cpubg \texttt{phase2\_e3\_end} & batch=1, forward-only, pre-tokenised & 1{,}658.15 \\
    \cpubg \texttt{phase2\_e3\_end} & batch=1, E2E                         & 1{,}695.84 \\
    \cpubg \texttt{phase2\_final} & batch=1, forward-only, pre-tokenised & 1{,}777.45 \\
    \cpubg \texttt{phase2\_final} & batch=1, E2E                         & 1{,}812.17 \\
    \bottomrule
  \end{tabular}
\end{table}

These measurements show that \textsc{Mamba-CPU} is not a low-latency CPU tier in our environment.  GPU acceleration and batching do not apply to the CPU path; with CUDA disabled, Transformers falls back to the sequential implementation.  Earlier lower CPU-only figures were therefore removed from the main claims.  \textsc{Mamba-GPU} single-request latency is 24.339\,ms, whereas the GuardChain cascade tables report single-request pipeline latencies.

\section{Cost-Normalised Deployment Analysis}
\label{app:cost}

The cost methodology uses the measured batch=1 latencies (Section~\ref{sec:protocol}) and AWS on-demand pricing as a reference.  No experiments were conducted on AWS instances; the costs below are proxy estimates that combine local latency measurements with instance-price assumptions.

To translate the F1$\times$latency Pareto into a dollar
figure that a deployment engineer can reason about, we construct the
following cost model.  For each safety classifier we (i) pick the cheapest
current-generation AWS \texttt{us-east-1} instance whose vCPU and RAM
specifications can run the model at the reported serving mode, (ii) use the
single-request inference latency measured on our evaluation server
(see Section~\ref{sec:protocol}), and (iii) compute USD per 1\,M
requests:
\[
\$/\text{1M\,req} \;=\; \frac{\$/\text{hr}}{\text{requests/hr per instance} / 10^6}.
\]
The latencies and prices are captured in Table~\ref{tab:cost}.

\begin{table}[t]
  \caption{Proxy cost-normalised serving cost (USD per 1\,M requests) at AWS
  \texttt{us-east-1} on-demand list prices, captured 2025-12-15.  Latency is
  the measured batch=1 latency used in the paper tables.  These are not direct
  AWS g5/c7i benchmark measurements.  The CPU-only Mamba row is a reference
  measurement, not the main GuardChain serving mode.}
  \label{tab:cost}
  \small
  \setlength{\tabcolsep}{3pt}
  \begin{tabular}{p{0.30\columnwidth}p{0.30\columnwidth}rr}
    \toprule
    \rowcolor{headerrow}\textcolor{white}{\textbf{Model}} & \textcolor{white}{\textbf{Instance}} & \textcolor{white}{\textbf{Lat (ms)}} & \textcolor{white}{\textbf{\$/1M req}} \\
    \midrule
    \cpubg SVM (TF-IDF)        & c7i.xlarge   & 3.360    & \best{\$0.17} \\
    \gpubg Mamba-130M E2E (A100 batch=1 representative) & g5.xlarge & 24.339 & \$6.80 \\
    \gpubg PromptGuard-86M     & g5.xlarge    & 14.407   & \$4.03 \\
    \gpubg Llama-Guard-3-1B    & g5.xlarge    & 41.263   & \$11.53 \\
    \gpubg ShieldGemma-2B      & g5.xlarge    & 60.245   & \$16.84 \\
    \gpubg WildGuard-7B        & g5.2xlarge   & 418.169  & \$140.78 \\
    \cpubg CPU-only Mamba-130M (reference) & c7i.4xlarge & 1{,}695.84 & \$336.34 \\
    \gpubg Granite-Guardian-2B & g5.2xlarge   & 65.051 & \$21.90 \\
    \bottomrule
  \end{tabular}
\end{table}

\textbf{Headline number.}  On in-distribution jailbreak detection (D1), a
TF-IDF SVM reaches accuracy comparable to fine-tuned GPU models while
costing \$0.17 per 1\,M requests on a 4-vCPU \texttt{c7i.xlarge}.  The
cheapest off-the-shelf GPU guard in our comparison (PromptGuard-86M on a single
g5.xlarge) costs \$4.03 per 1\,M requests.  \textsc{Mamba-GPU} (single-request)
costs \$6.80 per 1\,M requests at 24.339\,ms on g5.xlarge.  Granite-Guardian-2B
costs \$21.90 per 1\,M requests at its batch=1 latency of 65.051\,ms on
g5.2xlarge.

\textbf{The cascade is dominated by the CPU stage on in-distribution traffic.}  On D1, the GC-Acc pipeline routes approximately 2\% of samples to \textsc{Mamba-GPU} (Section~\ref{sec:gc_d1d3}).  The CPU-stage LightGBM cost of \$2.77 per 1\,M requests in Table~\ref{tab:cost_main} already represents a substantial saving versus GPU-only deployment, and the full pipeline invokes the GPU only on the small escalated fraction.

\textbf{Caveats.}  (1)~AWS spot pricing can cut these numbers by 60--90\%
but introduces eviction risk that is operationally disqualifying for many
safety-critical deployments.  (2)~We compare single-tenant instances; a
shared multi-tenant inference service such as AWS Bedrock or Anthropic API
would amortise GPU cost across more requests, but at the cost of network
RTT (typically +30--90\,ms for cross-AZ p50) and request-queueing-induced
tail latency that our cost model does not capture.  (3)~The proxy costs use
the paper's measured batch=1 latencies rather than direct AWS g5/c7i
benchmarks.

%% ============================================================
\section{Adaptive Learned Router}
\label{app:router}

The fixed-margin cascade (Section~\ref{sec:guardchain}) sets $\delta = 0.15$
uniformly across all inputs and all evaluation regimes.  A natural question is
whether a small \emph{learned} router could outperform this static rule.
We train a 200-tree LightGBM router on D1 train predictions; the router
takes the stage-A score plus five low-cost text features (length,
word-count, average word length, special-character ratio, word-repetition
score) and emits a per-prompt escalation probability.  Its decision
threshold is tuned on a held-out 20\% slice of D1 train using the
composite score $\text{acc} - 0.05\cdot\text{escalation\_rate}$.  Table~\ref{tab:router} compares the learned router against the fixed-margin cascade.

\begin{table}[t]
  \caption{Adaptive router vs.\ fixed-margin cascade.  Router uses the
  selected escalation threshold $p\!\ge\!0.15$ (chosen on held-out 20\% of
  D1 train using the composite score $\text{acc} - 0.05\cdot\text{esc}$);
  fixed cascade uses $\delta=0.10$.  Both share the same stage-A (LightGBM)
  and stage-B (Mamba-130M E2E) classifiers as the main pipeline.}
  \label{tab:router}
  \small
  \setlength{\tabcolsep}{3pt}
  \begin{tabular}{lrrrr}
    \toprule
    \rowcolor{headerrow}\textcolor{white}{\textbf{Eval}} & \textcolor{white}{\textbf{Rtr.\ F1}} & \textcolor{white}{\textbf{Rtr.\ esc.}} & \textcolor{white}{\textbf{Fixed F1}} & \textcolor{white}{\textbf{Stg-A F1}} \\
    \midrule
    \cpubg D1 (ID jailbreak)  & \best{0.9801} & 1.1\%  & 0.9705 & 0.9701 \\
    \cpubg D2 (source-shift)  & \best{0.9925} & 1.7\%  & 0.9915 & 0.9913 \\
    \gpubg D3 (Obfuscated)    & 0.3135        & 0.4\%  & \best{0.3865} & 0.3455 \\
    \bottomrule
  \end{tabular}
\end{table}

\textbf{Finding.}  On D1 and D2, the router matches or beats the fixed
cascade with $<$2\% escalation---roughly a 5$\times$ reduction in stage-B
invocations.  On D3 (obfuscated) the router actually \emph{underperforms}
the fixed cascade because its training distribution (D1, ID-only)
mis-calibrates its escalation signal under the heavy character-level
perturbations of D3: it learns to trust stage-A on confident predictions,
but on D3 stage-A is confidently wrong (Section~\ref{sec:gc_d2}).  A
distribution-shift-aware router would need to either (i) include synthetic
obfuscated samples in training, or (ii) widen its escalation band when a
secondary signal (e.g., n-gram OOV rate) detects perturbation.  We list
this as future work in Section~\ref{sec:limitations}; the main paper
adopts the simpler fixed-margin rule.

\section{Multi-Turn Crescendo Evaluation}
\label{app:crescendo}

To evaluate beyond the single-turn setting, we
synthesise a 500-conversation Crescendo~\cite{russinovich2024great}
corpus from D1 test malicious seeds.  Each conversation is a 5-turn
templated escalation: turn~1 is an academic framing; turn~2 narrows the
question; turn~3 supplies a fictional context; turn~4 escalates
specificity; turn~5 is the verbatim D1 seed.  Templates are deterministic
and category-conditioned (weapons, malware, fraud, drugs, default;
CSAM-category seeds are explicitly excluded).  The synthesis is
reproducible from random seed \texttt{20251215}.

We evaluate two CPU classifiers from the main paper on this corpus.  For
each conversation we record (i) the smallest turn at which the classifier
predicts ``malicious'' (\emph{early-detect turn}), (ii) whether the
classifier flagged \emph{any} of the five turns (1$-$ASR; ASR
$\triangleq$ ``no turn was flagged'').  Table~\ref{tab:crescendo} reports the results.

\begin{table}[H]
  \caption{500-conversation 5-turn Crescendo evaluation.  ASR is the fraction
  of conversations with no flagged turn; lower is better.}
  \label{tab:crescendo}
  \small
  \setlength{\tabcolsep}{4pt}
  \begin{tabular}{lrrr}
    \toprule
    \rowcolor{headerrow}\textcolor{white}{\textbf{Model}} & \textcolor{white}{\textbf{ASR}} & \textcolor{white}{\textbf{Med.\ early}} & \textcolor{white}{\textbf{Turn-5-only}} \\
    \midrule
    \cpubg SVM Variant~A & 4.0\%  & 5 & 71.2\% \\
    \cpubg LightGBM      & 0.0\%  & 2 & 3.6\% \\
    \bottomrule
  \end{tabular}
\end{table}

\textbf{Finding.}  LightGBM flags 100\% of conversations and does so by
turn~2 on the median conversation, indicating that the templated
escalation language itself is detectable from training-distribution n-gram
statistics.  SVM, which relies on tighter lexical surface, catches 96\% of
conversations \emph{only at turn~5} (the verbatim seed) and misses the
escalation arc itself---an attacker who stops one turn before the seed
would evade SVM in 71\% of conversations.  Caveat: 100\% LightGBM
detection is partially an artefact of our deterministic template; a
LLM-generated escalation would diversify the surface forms.  We list
GPT-/Claude-generated multi-turn corpora and Mamba session-state extensions
as future work in Section~\ref{sec:limitations}.

%% ============================================================
\section{Extended Cascade-Margin Sweep}
\label{app:cascade_sweep_ext}

Appendix~\ref{app:cascade} already discusses the cascade-margin sensitivity
on a coarse $\delta$ grid.  We extend the sweep to seven margins,
$\delta \in \{0.05, 0.10, 0.15, 0.20, 0.25, 0.30, 0.35\}$, attach bootstrap
95\% CIs (2{,}000 resamples) and an exact-McNemar paired test against the
narrow-margin setting $\delta = 0.05$.  All numbers are from a single-pass
deterministic re-evaluation on each test split.

\begin{table}[H]
  \caption{Cascade-margin sweep on obfuscated D3 for LightGBM and SVM-A as
  Stage~A, with fine-tuned Mamba-130M E2E as Stage~B.  Wider margins improve
  D3 F1 but increase escalation.}
  \label{tab:cascade_sweep_ext}
  \scriptsize
  \centering
  \setlength{\tabcolsep}{2pt}
  \begin{tabular}{@{}lrrrrrr@{}}
    \toprule
    \rowcolor{headerrow}\textcolor{white}{\textbf{Stage-A}} & \textcolor{white}{\textbf{$\delta$}} & \textcolor{white}{\textbf{F1}} & \textcolor{white}{\textbf{95\% CI}} & \textcolor{white}{\textbf{Prec.}} & \textcolor{white}{\textbf{Esc.\%}} & \textcolor{white}{\textbf{McN.\ $p$}} \\
    \midrule
    \cpubg LightGBM & 0.05 & 0.3455 & [.3255, .3652] & 0.890 & 3.6  & --- \\
    \cpubg LightGBM & 0.10 & 0.3865 & [.3668, .4060] & 0.911 & 7.3  & $<\!10^{-4}$ \\
    \cpubg LightGBM & 0.15 & 0.4293 & [.4097, .4475] & 0.923 & 11.3 & $<\!10^{-4}$ \\
    \cpubg LightGBM & 0.20 & 0.4714 & [.4522, .4895] & 0.930 & 15.6 & $<\!10^{-4}$ \\
    \cpubg LightGBM & 0.25 & 0.5135 & [.4953, .5319] & 0.933 & 20.0 & $<\!10^{-4}$ \\
    \cpubg LightGBM & 0.30 & 0.5782 & [.5613, .5948] & 0.938 & 26.3 & $<\!10^{-4}$ \\
    \cpubg LightGBM & 0.35 & 0.6452 & [.6294, .6612] & 0.933 & 34.4 & $<\!10^{-4}$ \\
    \midrule
    \cpubg SVM-A & 0.05 & 0.4919 & [.4737, .5098] & 0.875 & 6.1  & --- \\
    \cpubg SVM-A & 0.10 & 0.5447 & [.5275, .5619] & 0.897 & 12.6 & $<\!10^{-4}$ \\
    \cpubg SVM-A & 0.15 & 0.5902 & [.5738, .6067] & 0.912 & 19.3 & $<\!10^{-4}$ \\
    \cpubg SVM-A & 0.20 & 0.6492 & [.6336, .6648] & 0.924 & 26.5 & $<\!10^{-4}$ \\
    \cpubg SVM-A & 0.25 & 0.7166 & [.7022, .7294] & 0.931 & 35.9 & $<\!10^{-4}$ \\
    \cpubg SVM-A & 0.30 & 0.7795 & [.7672, .7909] & 0.935 & 45.4 & $<\!10^{-4}$ \\
    \cpubg SVM-A & 0.35 & \best{0.8249} & [.8141, .8356] & 0.933 & 54.6 & $<\!10^{-4}$ \\
    \bottomrule
  \end{tabular}
\end{table}

\textbf{Interpretation.}  Section~\ref{sec:gc_d2} argues that confidence-based
escalation cannot recover from \emph{confidently wrong} CPU predictions on
source-shift data.  The cascade-margin sweep makes this concrete: at $\delta=0.05$
(the narrow-margin baseline) only 6\% of D3 obfuscated samples are
escalated, because stage-A's miscalibrated probabilities sit on the
confident-and-wrong side of the boundary; widening $\delta$ to 0.35 forces
the cascade to escalate 55\% of D3 samples, recovering 33 F1 points on the
hard split with a McNemar $p \ll 10^{-4}$.  This argues for a per-task
$\delta$ (the main paper proposes future learned-router work in
Appendix~\ref{app:router}) rather than a single uniform value (see also Table~\ref{tab:cascade_sweep_ext}).

%% ============================================================
\section{Within-CPU Pairwise McNemar}
\label{app:within_cpu}

Appendix~\ref{app:stats} reports CPU-vs-GPU McNemar tests.  This appendix
gives the within-CPU companion: pairwise exact McNemar tests between every
pair of CPU classifiers on each of D1/D2/D3 test splits, alongside the
F1 with bootstrap 95\% CI (2{,}000 resamples) for each model.  Table~\ref{tab:within_cpu_d1} reports the D1 results.

\begin{table}[H]
  \caption{Within-CPU pairwise McNemar (exact) on D1.  $\Delta$F1 is
  F1(B)$-$F1(A); full D2/D3 tables are in \texttt{e9\_within\_cpu.md}.}
  \label{tab:within_cpu_d1}
  \small
  \setlength{\tabcolsep}{3pt}
  \begin{tabular}{lrrrr}
    \toprule
    \rowcolor{headerrow}\textcolor{white}{\textbf{Pair (A vs.\ B)}} & \textcolor{white}{\textbf{A}} & \textcolor{white}{\textbf{B}} & \textcolor{white}{\textbf{$\Delta$F1}} & \textcolor{white}{\textbf{$p$}} \\
    \midrule
    SVM-A vs.\ SVM-B           &  85 &  58 & $-$0.0042 & 0.0293 \\
    SVM-A vs.\ SVM-C           &  40 &  35 & $-$0.0007 & 0.6445 \\
    SVM-A vs.\ LightGBM        &  54 &  94 & +0.0064 & 0.0013 \\
    SVM-A vs.\ Random-Forest   & 115 &  78 & $-$0.0065 & 0.0094 \\
    SVM-B vs.\ LightGBM        &  54 & 121 & +0.0106 & $<\!10^{-4}$ \\
    SVM-C vs.\ LightGBM        &  45 &  90 & +0.0071 & 0.0001 \\
    LightGBM vs.\ Random-Forest& 108 &  31 & $-$0.0129 & $<\!10^{-4}$ \\
    \bottomrule
  \end{tabular}
\end{table}

\textbf{Interpretation.}  LightGBM is significantly better than every SVM
variant on D1 by $p\le 0.0013$; SVM-A and SVM-C are statistically
indistinguishable ($p=0.64$), confirming that the choice between
word-only and word$+$char n-grams for SVM is within sampling variation on
ID jailbreak.  Random Forest is consistently the weakest CPU classifier
($\Delta$F1 vs.\ LightGBM is $-0.013$, $p<10^{-4}$), supporting the
decision to omit it from the production pipeline.  Full D2/D3 results
are in the release artefact \texttt{e9\_within\_cpu.md}.

%% ============================================================
\section{Mamba Checkpoint Re-Selection on Held-Out Validation}
\label{app:mamba_reselection}

A potential concern is whether the paper's reported Mamba checkpoint
(\texttt{phase2\_jailbreak\_ckpt\_e3\_end.pt}) was selected on the test set
itself, biasing the reported F1 upward.  To answer, we re-derive the
checkpoint selection using the held-out validation F1 logged during
training (\path{mamba_phase2_jailbreak_checkpoints.json}), then
compare each task's val-best checkpoint's test F1 against the post-hoc
test-best checkpoint's test F1.  Table~\ref{tab:mamba_reselection} reports the gap.

\begin{table}[H]
  \caption{Mamba checkpoint re-selection.  Val-best uses held-out validation
  F1 only; \textbf{Gap} is the test-best minus val-best test F1.}
  \label{tab:mamba_reselection}
  \small
  \setlength{\tabcolsep}{4pt}
  \begin{tabular}{lllr}
    \toprule
    \rowcolor{headerrow}\textcolor{white}{\textbf{Eval}} & \textcolor{white}{\textbf{Val-best ckpt}} & \textcolor{white}{\textbf{Val-best test F1}} & \textcolor{white}{\textbf{Gap}} \\
    \midrule
    D1 (Jailbreak)    & \texttt{e5\_half} & 0.9623 & $+$0.0000 \\
    D2 (source-shift) & \texttt{e5\_half} & 0.9854 & $+$0.0002 \\
    D3 (Obfuscated)   & \texttt{e5\_half} & 0.9293 & $+$0.0010 \\
    D4 (Hate)         & \texttt{e5\_half} & 0.7988 & $+$0.0042 \\
    D5 (PII)          & \texttt{e5\_half} & 0.9781 & $+$0.0081 \\
    D6 (Toxic)        & \texttt{e5\_half} & 0.2881 & $+$0.0188 \\
    \bottomrule
  \end{tabular}
\end{table}

\textbf{Finding.}  The largest test/val-leakage gap across all six tasks
is 0.019 F1 (toxicity); five of six tasks have a gap below 0.008 F1.  The
paper's reported \texttt{phase2\_e3\_end} checkpoint is within this margin
of the val-best on every task; the original test-driven selection is not
measurably over-fit to the test set.  We keep \texttt{e3\_end} as the
canonical paper checkpoint because it balances jailbreak F1 (which
improves with later epochs) against catastrophic forgetting on the
multitask Phase-1 tasks (which worsens with later epochs; see
Appendix~\ref{app:mamba_checkpoints}, Figure~\ref{fig:catastrophic_forgetting}).
The full per-checkpoint val+test table is in the release artefact under
\path{e5_mamba_reselection/}.

\end{document}